\documentclass[lettersize,journal]{IEEEtran}

\usepackage{xcolor}

\usepackage{xurl}

\usepackage{booktabs}
\usepackage[ruled]{algorithm2e}
\usepackage{algpseudocode}
\usepackage{enumerate}
\usepackage[english]{babel}
\usepackage{blindtext}
\usepackage{amsmath}

\usepackage{amssymb}
\usepackage{xspace}
\usepackage{cite}
\usepackage{multirow}
\usepackage{threeparttable}
\usepackage{float}
\usepackage{flafter}
\usepackage{comment}
\usepackage[skip=0pt]{caption}
\usepackage{subcaption}
\usepackage{graphicx}
\usepackage{ulem}
\usepackage{verbatim}
\usepackage{pifont}
\usepackage{lipsum}
\usepackage{hyperref}

\def\fig{Fig.\xspace}

\def\tab{Tab.\xspace}

\def\ie{{\textit{i.e.}\xspace}} 
\def\eg{{\textit{e.g.}\xspace}}

\def\etc{{\textit{etc}\xspace}}

\newcommand{\head}[1]{{\noindent \textbf{#1:}}}

\graphicspath{{./figures/}}
\graphicspath{{./pdf/}}
\DeclareGraphicsExtensions{.pdf,.jpeg,.png}

\ifodd 0
\newcommand{\com}[1]{\textbf{\color{red}(COMMENT: #1)}} %
\newcommand{\todo}[1]{\textbf{{\color{orange}(TODO: #1)}}}
\newcommand{\unused}[1]{{\color{gray}#1}}
\newcommand{\sheng}[1]{\textbf{\color{olive}(Sheng: #1)}} %
\else

\newcommand{\com}[1]{}
\newcommand{\todo}[1]{}
\newcommand{\unused}[1]{}
\newcommand{\sheng}[1]{}

\fi

\ifodd 0
\newcommand{\tmc}[1]{{\color{blue}#1}}
\else
\newcommand{\tmc}[1]{#1}
\fi

\newif\ifcolorful
\colorfulfalse %

\newcommand{\colortable}[1]{%
    \ifcolorful
      {\color{blue}#1}
    \else
      #1
    \fi
}

\def\sysname{\textit{CardioLive}\xspace}

\begin{document}
\title{\sysname: Empowering Video Streaming  with Online Cardiac Monitoring
}
\author{
\IEEEauthorblockN{
Sheng Lyu\textsuperscript{1}, 
Ruiming Huang\textsuperscript{1}, 
Sijie Ji\textsuperscript{1},
Yasar Abbas Ur Rehman\textsuperscript{2},
Lan Ma\textsuperscript{2},
Chenshu Wu\textsuperscript{1}
}
\thanks{
Contact Email: Sheng Lyu   \href{mailto:shenglyu@connect.hku.hk}{shenglyu@connect.hku.hk}, Chenshu Wu \href{mailto:chenshu@cs.hku.hk}{chenshu@cs.hku.hk}. Chenshu Wu is the corresponding author. Ruiming Huang did this work when he was a research assistant at HKU. Sijie Ji participated in this work when she was a postdoctoral fellow at HKU.
}

  \IEEEauthorblockA{\textsuperscript{1}Department of Computer Science, The University of Hong Kong, Hong Kong SAR}
  
  \IEEEauthorblockA{\textsuperscript{2}TCL AI Lab, Hong Kong SAR} 
}

\maketitle
\begin{abstract}
Online Cardiac Monitoring (OCM) emerges as a compelling enhancement for the next-generation video streaming platforms. It enables various applications, including remote health, affective computing, and deepfake detection. 
Yet the physiological information encapsulated in the video streams has long been neglected. In this paper, we present the design and implementation of \textit{CardioLive}\xspace, the first online cardiac monitoring system in video streaming platforms. 
We leverage the naturally co-existing video and audio streams and devise \texttt{CardioNet}, the first audio-visual network to learn the cardiac series. It incorporates multiple unique designs to extract temporal and spectral features, ensuring robust performance under realistic streaming conditions. To enable the Service-On-Demand OCM, we implement \textit{CardioLive}\xspace as a plug-and-play middleware service and develop systematic solutions to practical issues including changing FPS and unsynchronized streams. 
Extensive evaluations demonstrate the effectiveness of our system. We achieve a Mean Squared Error of 1.79 BPM error, outperforming the video-only and audio-only solutions by 69.2\% and 81.2\%, respectively. \textit{CardioLive}\xspace achieves average throughput of 115.97 and 98.16 FPS in Zoom and YouTube. We believe our work opens up new applications for video stream systems. Code is available at \href{https://github.com/aiot-lab/CardioLive}{https://github.com/aiot-lab/CardioLive}.
\end{abstract}

\begin{IEEEkeywords}
Mobile Computing Systems, Audio-Visual Learning, Middleware, Vital-Signs, Multimodal Sensing.
\end{IEEEkeywords}

\begin{figure}[t]
    \centering
    \includegraphics[width=\linewidth]{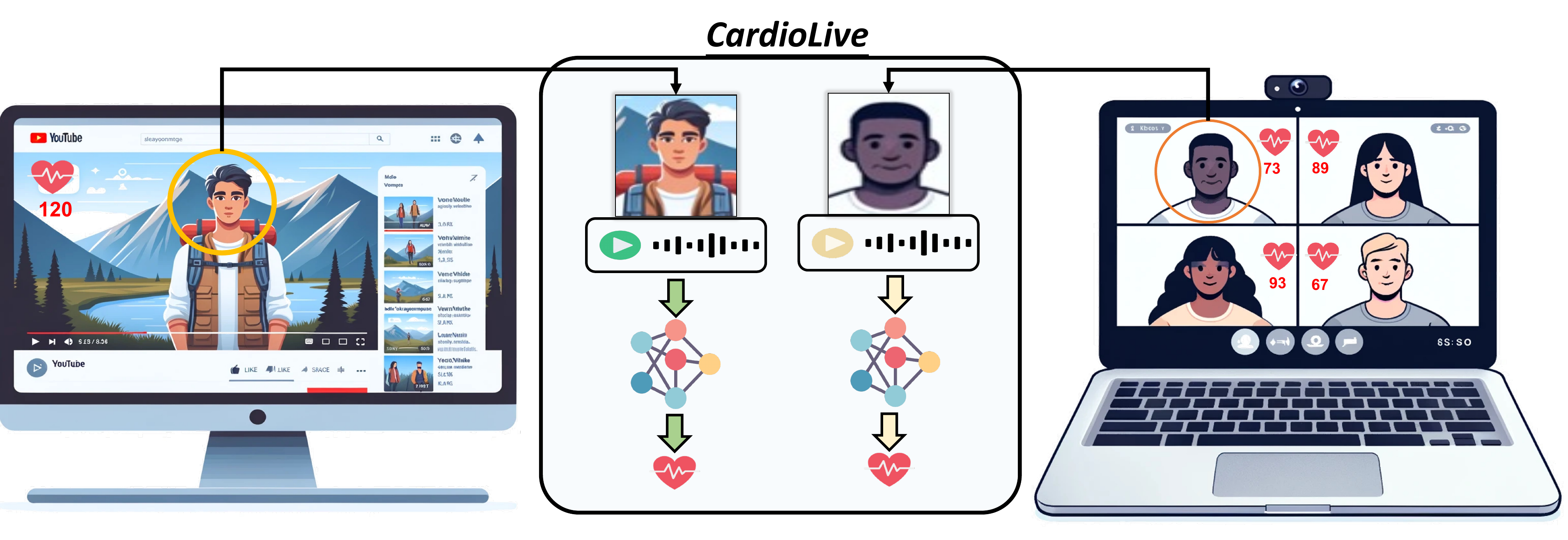}
    \caption{Online Cardiac Monitoring (OCM).}
    \label{fig:intro}
\end{figure}

\section{Introduction}

Video streaming has exploded in recent years, with no slowdown in sight. 
From TikTok that have turned live video sharing into a global phenomenon, to Zoom, which has become synonymous with remote work, video streaming has woven itself into the fabric of our daily lives. 
The market is booming steadily \cite{VideoStreamingSVoD}, reflecting our collective appetite for real-time, interactive, and accessible content.

Online Cardiac Monitoring (OCM) can be one intriguing enhancement for the next-generation video streaming platforms. 
The rich tapestry of video and audio in streaming not only provides the context of actions, movement, and human activities, \etc, but it also embeds subtle cardiac events, which have long been neglected in contemporary multimedia systems. 
Uncovering such physiological information would bring various benefits. For remote health, physicians could remotely access real-time cardiac data without the need for specialized equipment \cite{HeartDiseaseSymptoms}. Similarly, in video gaming, displaying a player's heart rate during live streams could add additional excitement and engagement for viewers \cite{PulsoidRealtimeHearta}. 
OCM also plays a pivotal role in online conferences or interviews, where emotional responses inferred from cardiac data \cite{wang2020you, sun2022estimating} could enrich interactions, making them more nuanced and meaningful. 
Furthermore, the potential for this technology extends into security and fraud detection against digital impersonation techniques like deepfakes \cite{liu2019cardiocam, qi2020deeprhythm}. 
These applications of OCM underscore its potential to revolutionize video streaming, making it not just a tool for communication and entertainment but also a platform for health monitoring, affective computing, emotional intelligence, and security.

However, existing OCM either relies on specified hardware (\eg, heartbeat belt, wearables) or introduces additional modalities \cite{liu2015tracking, yang2016monitoring, chen2021movi, xue2025ppg}.
These approaches suffer from extra cost and are often misaligned with live streams. Moreover, sensing-based approaches necessitate active probing signals \cite{wang2023df, qian2018acousticcardiogram, lyu2024ase}, which are impractical in streaming. 
A video streaming system that seamlessly enables OCM in pervasive contexts without additional hardware still lacks.

In this paper, we ask: \textit{Can we incorporate accurate and robust online cardiac monitoring into a video-streaming system without introducing additional hardware or modalities?} 
To build such a system, we answer the following key questions:

First, \textit{what information should we take from the video streaming system to monitor the cardiac activities?} 
Existing works \cite{chen2018deepphys, liu2020multi, niu2020video, li2023learning, yu2019remote,yu2019remoteCompress, yu2023physformer++, liu2023efficientphys, zou2024rhythmformer} on extracting heart rate from human faces focus on remote photoplethysmography (rPPG), which leverages solely video. These video-only solutions suffer from low illumination conditions, head movement, and orientation. 
Recent progress in cardiac vocal interfaces \cite{xu2022hearing} inspires us to infer heart rate from human speech. However, audio signals are usually sensitive to noise interference and lack contextual background information, rendering them less robust in real-life scenarios and requiring user calibration. 
Conceptually, video provides detailed visual context while sound exhibits resilience to varying light conditions and body motions. 
Consequently, they offer complementary advantages to enhance cardiac monitoring. This motivates us to move beyond video-only or audio-only solutions and investigate new designs to combine the naturally co-existing video and audio streams. 

Second, \textit{how to tackle real-world problems to make this system robust and accurate?} Unveiling the cardiac activity from video and audio is challenging. The information is easy to be overshadowed by more prominent body movements, environmental dynamics, and/or ambient noise. Previous works \cite{liu2020multi, niu2020video, li2023learning, yu2023physformer++, liu2023efficientphys, zou2024rhythmformer} primarily evaluate models on well-controlled datasets featuring static subjects under optimized light conditions and viewing angles, which simplifies the problems yet becomes unrealistic in real-world settings. 
The task gets even more challenging when deployed in live video streaming environments, due to the discrepancies in frame rates and degraded image quality.
To deliver an accurate and robust system in practice, novel techniques are desired to effectively discern subtle cardiac signals amidst various disturbances while combating fluctuating frame rates and drifted misalignment of the streams.

Third, \textit{how to enable Service-On-Demand (SoD) cardiac monitoring in video streaming system} Despite the promise of the integration,
enabling SoD for users poses significant challenges due to the complexity of modern video streaming systems. These platforms vary widely, encompassing formats such as conferences \cite{SkypeStayConnected, TeamsChannelsMicrosoft, OnePlatformConnect}, Video-On-Demand (VoD) \cite{StreamTVMovies, NetflixSingaporeWatch}, live streaming \cite{YouTube, ExploreFindYour}, \etc, each with its own technical and operational nuances. 
These providers balance the demands of real-time data processing with the need for immediate accessibility and minimal latency while not interfering with the original streams. At the same time, deploying our service on edge (\eg, browsers) benefits from preserving privacy, while getting access to the data yields another challenge. One naive way is to deploy our models over the WebRTC peers, but it lacks scalability and versatility. To this end, we are motivated to establish a plug-and-play service that can be seamlessly integrated into video streaming systems, whether hosted on servers or edges.

In this paper, we present \sysname, the first-of-its-kind OCM system, that can continuously infer the heart rate in video streaming systems. At the core of \sysname, we design a novel audio-video network, \texttt{CardioNet}, that effectively learns the nuanced cardiac activities from facial regions and human voices. 
We further devise systematic solutions to deploy \sysname as a middleware service to support the SoD online cardiac monitoring. 
We introduce practical techniques to handle changing FPS and unsynchronized streams. 
Through in-depth analyses of the streaming architectures, we design effective data hooks and a novel packet buffer, which can be easily integrated with various video streaming systems.

Extensive experiments have been done to validate the effectiveness of \sysname. 
Our evaluation results show that \sysname achieves a mean absolute error (MAE) of 1.79 BPM and root mean square error (RMSE) of 3.25 BPM, largely outperforming the video-only solutions by 69.2\% in MAE and 61.4\% in RMSE,  and the audio-only solution by 81.2\% in MAE and 76.8\% in RMSE. 
As for \sysname service, we implement our system on two ends, a meeting platform (Zoom) and a content provider (YouTube), respectively. We achieve the overall throughput of 115.97 FPS and 98.16 FPS for each platform, respectively, ensuring smooth updates without disrupting the original streams.
These results highlight the robustness and accuracy of \sysname, confirming its potential for widespread application in video streaming systems.

\head{Contributions} We conclude our contributions as follows:

\noindent\ding{182} To the best of our knowledge, we are the first to combine video and audio for cardiac monitoring in video streaming systems. Our solution outperforms video-only or audio-only approaches, especially under adverse conditions in practice.

\noindent\ding{183} We develop \texttt{CardioNet}, a novel audio-video pipeline that can uncover the nuanced heart rate. Our experiments validate the robustness against different conditions.

\noindent\ding{184} We implement \sysname as a service-based plug-and-play middleware that can seamlessly be integrated into mainstream platforms for real-time streaming.

\tmc{
\section{Design Scope}
In this section, we will discuss what potential benefits \sysname can bring about and the research scope of this paper.

\head{Application Momentum}
Consider a scenario where users on platforms such as Zoom or YouTube can access real-time cardiac monitoring. With just a single click, users see their heart rate, providing immediate insights into their emotional and physiological states, including what others are thinking about, whether they are in good health, and how exciting the game is. By online cardiac monitoring, these platforms could significantly enhance user engagement and interactivity. Particularly, \sysname can provide unique and compelling benefits in the downstream applications:

\noindent\ding{182} \textbf{Accessibility:} In many video streaming scenarios, such as live product demonstrations on TikTok or Zoom interviews, using wearables or additional hardware is often impractical. OCM can overcome this problem by leveraging modalities that already exist within video streams, thereby increasing accessibility for audiences and facilitating broader engagement.  It also promises wider dissemination of remote health, offering device-free cardiac monitoring compared to the latest work \cite{chen2024exploring} that relies on earphones. 

\noindent\ding{183} \textbf{Enhanced Analytical Abilities:} While there exist alternative approaches for tasks including affective computing \cite{mottelson2016affect, prajwal2023towards, wu2020emo, ahmad2024detecting} and deepfake detection \cite{yang2023avoid, demir2021deep}, the cardiac signal shows a strong correlation with them \cite{wascher2021heart, prajwal2023towards}, by capturing the subtle changes in heart rate. In this context, OCM provides an additional verification layer in a real-time and continuous manner, allowing experts to analyze behaviors. This analysis can help determine if someone is lying, happy, nervous, or engaging in deceptive behavior. 

\noindent\ding{184} \textbf{Entertainment:} Our work also presents a distinct chance for augmented entertainment. With the rise of live streaming, the audience can access the heart rates of celebrities, which opens up a new world for existing viewing experiences.

\begin{figure*}[t]
    \centering
    \includegraphics[width=\linewidth]{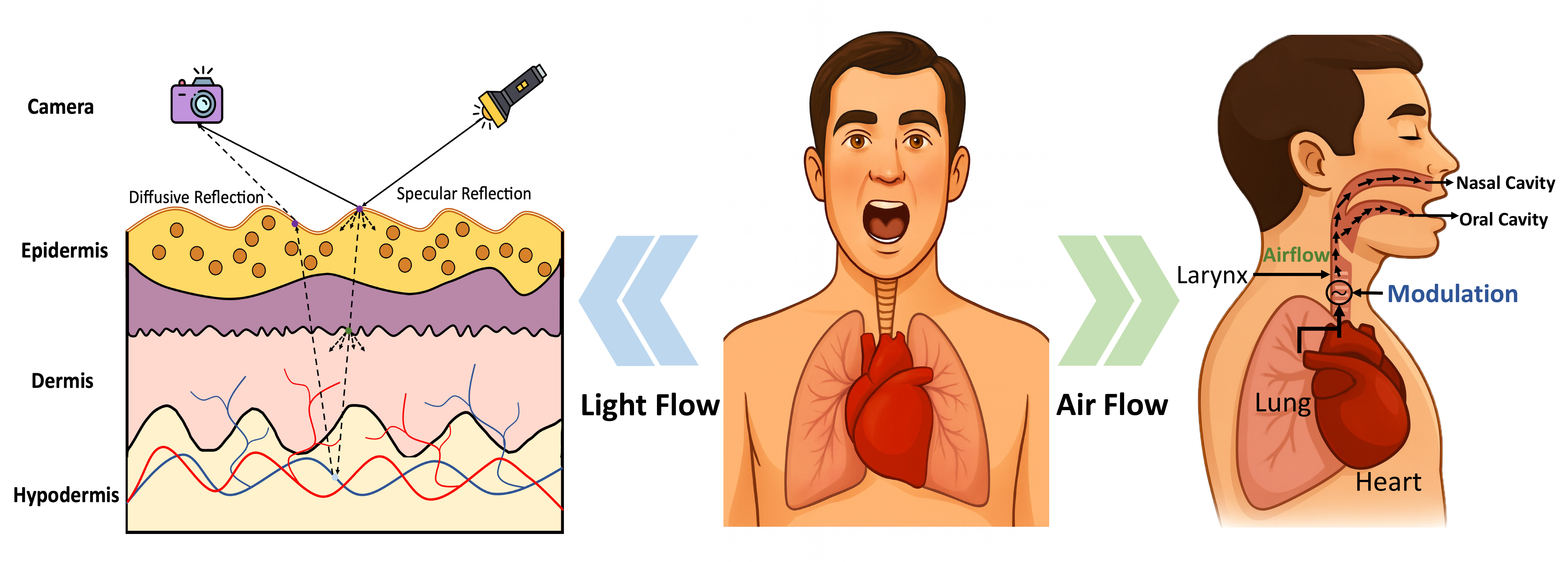}
    \caption{\tmc{Kinetics of Cardiac Learning.}}
    \label{fig:kinetics}
\end{figure*}

Despite the potential, there are \textit{no} existing solutions capable of achieving this integration without additional hardware. 
In this work, we focus on addressing this gap by leveraging the co-existence of audio and video signals, specifically in scenarios where a speaker is talking. This can be common in both entertainment and telehealth use cases. \textit{At the core of OCM is the accurate prediction of cardiac information.} Our system should robustly detect the heart rate from the video streaming systems by hooking the video and audio chunks. Once cardiac data is acquired, it can be further analyzed for various downstream tasks, including affective computing, remote health monitoring, and deepfake detection. Yet how cardiac monitoring is used for downstream tasks (\eg, emotions, lies, \etc) is not the focus of this paper. 

\head{Audio-Video Pair} We intend to integrate the video and audio information for cardiac monitoring. Leveraging the natural co-existence of audio and video modalities offers contemporary benefits as follows: 
\noindent \ding{182} Ubiquity: Video and audio streams are the most fundamental components in video streaming systems, while no additional hardware is needed.
\noindent \ding{183}  Feasibility: Both video and audio data contain the cardiac information (discussed in \S\ref{subsec: principle}).
\noindent \ding{184}  Complementarity: Audio and video offer different strengths and weaknesses. Audio is less interfered with by motion and light but is sensitive to noises. Video is more robust to noise but will fail in various body movements and non-optimized view angles. We will elaborate on the detailed analyses in \S\ref{sec: cardionet}.
We argue that in our primary target application scenarios—such as video conferencing, live streaming, and remote healthcare—human speech is inherently present alongside video. Our goal is to fully leverage the potential of these naturally coexisting signals. Additionally, our system is well-designed to seamlessly fall back to a video-only solution when audio quality degrades.

\head{\sysname as a service}
To deploy such an OCM system, a straightforward way is to build a self-hosted WebRTC service, which, however, does not scale to existing streaming systems. 
Therefore, for the sake of versatility, we establish a microservice to host \sysname for seamless integration with mainstream video streaming platforms.

\head{Privacy Concerns} Audio and video data are inherently sensitive and vulnerable to privacy breaches. However, in our proposed scenarios, privacy concerns are mitigated for several reasons. First, the primary purpose of audio and video data in this context is for communication. Therefore, participants are already receiving this data during the meetings, regardless of whether our system is activated or not. In other words, all participants have consented to share their audio and video within the video streaming applications, without requiring extra sensitive data inputs. Additionally, our system is implemented as a middleware solution within existing video streaming systems. These contemporary systems are subject to stringent privacy regulations. \sysname will operate in compliance with these established privacy frameworks.

In a nutshell, the audio-video pair appears to be an attractive choice for ubiquitous and practical OCM, yet it entails numerous challenges to build an accurate and robust multi-modal algorithm and system. We will present our model design in \S\ref{sec: cardionet} and leave the system implementation in \S\ref{sec: sys_design}.

}

\begin{figure*}[t]
\centering

\begin{subfigure}{\linewidth}
    \centering
    \begin{subfigure}{.32\linewidth}
        \includegraphics[width=\linewidth]{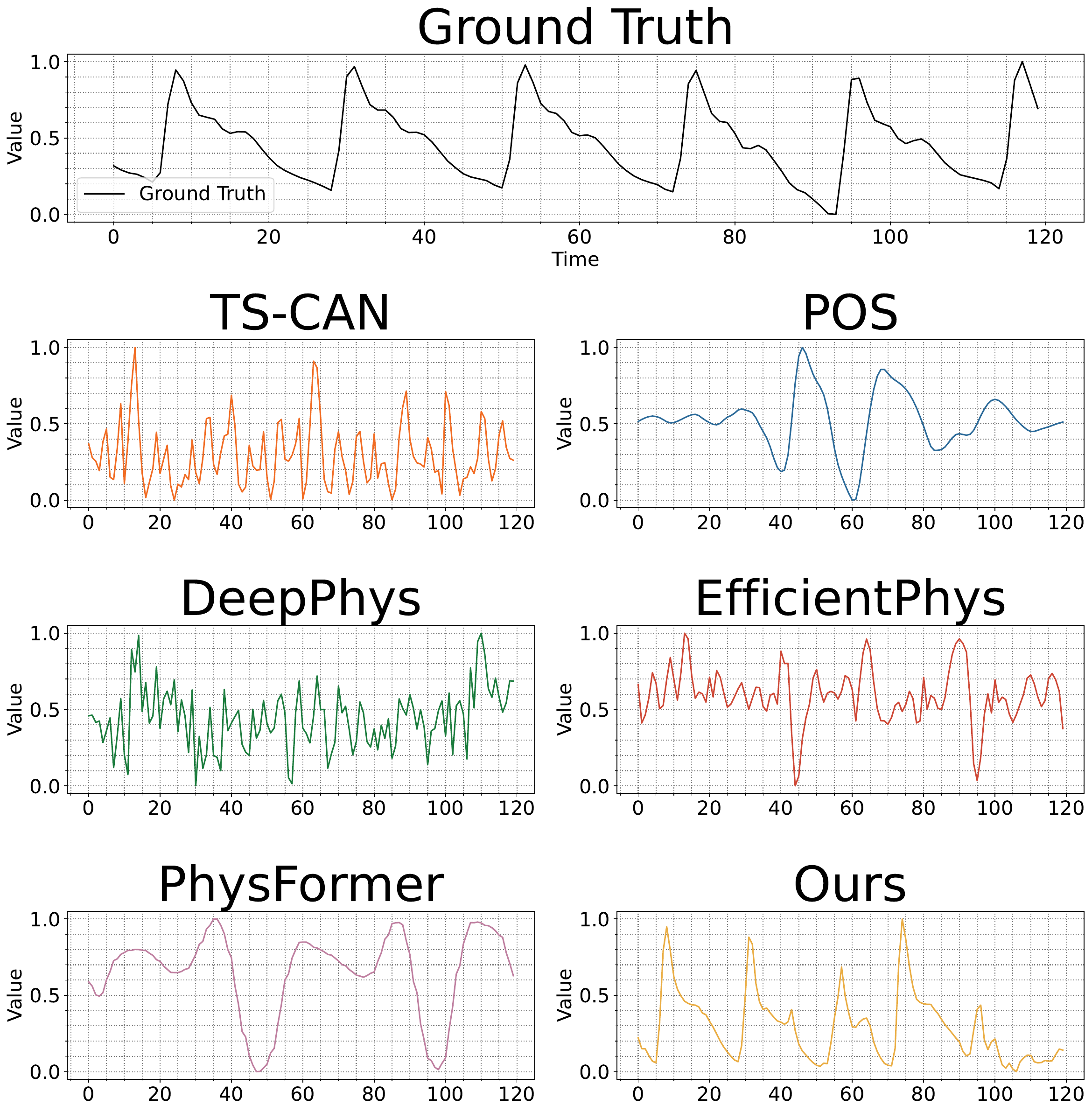}
        \caption{Stationary}
        \label{subfig:a}
    \end{subfigure}\hfill
    \begin{subfigure}{.32\linewidth}
        \includegraphics[width=\linewidth]{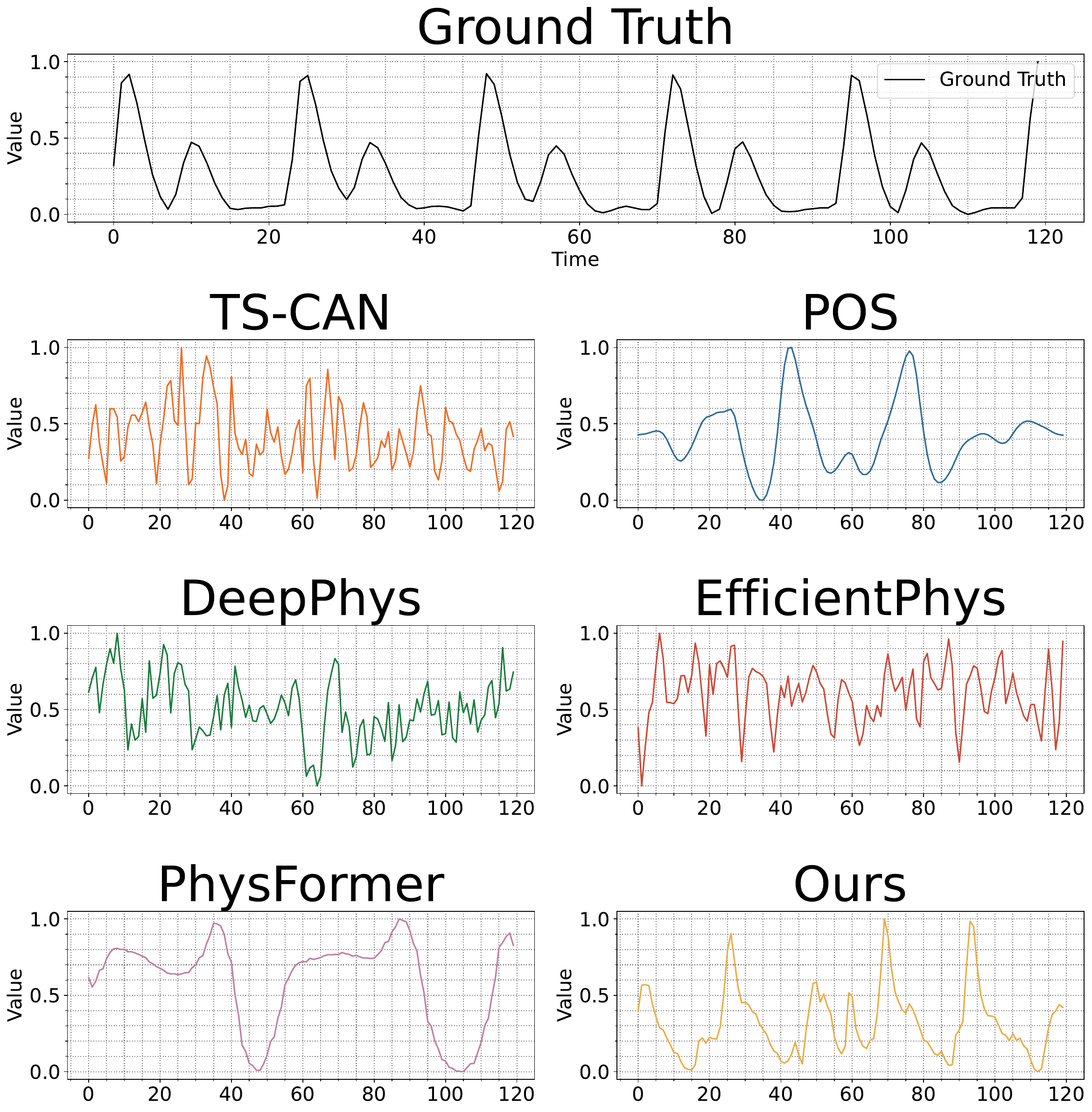}
        \caption{Rotation}
        \label{subfig:b}
    \end{subfigure}\hfill
    \begin{subfigure}{.32\linewidth}
        \includegraphics[width=\linewidth]{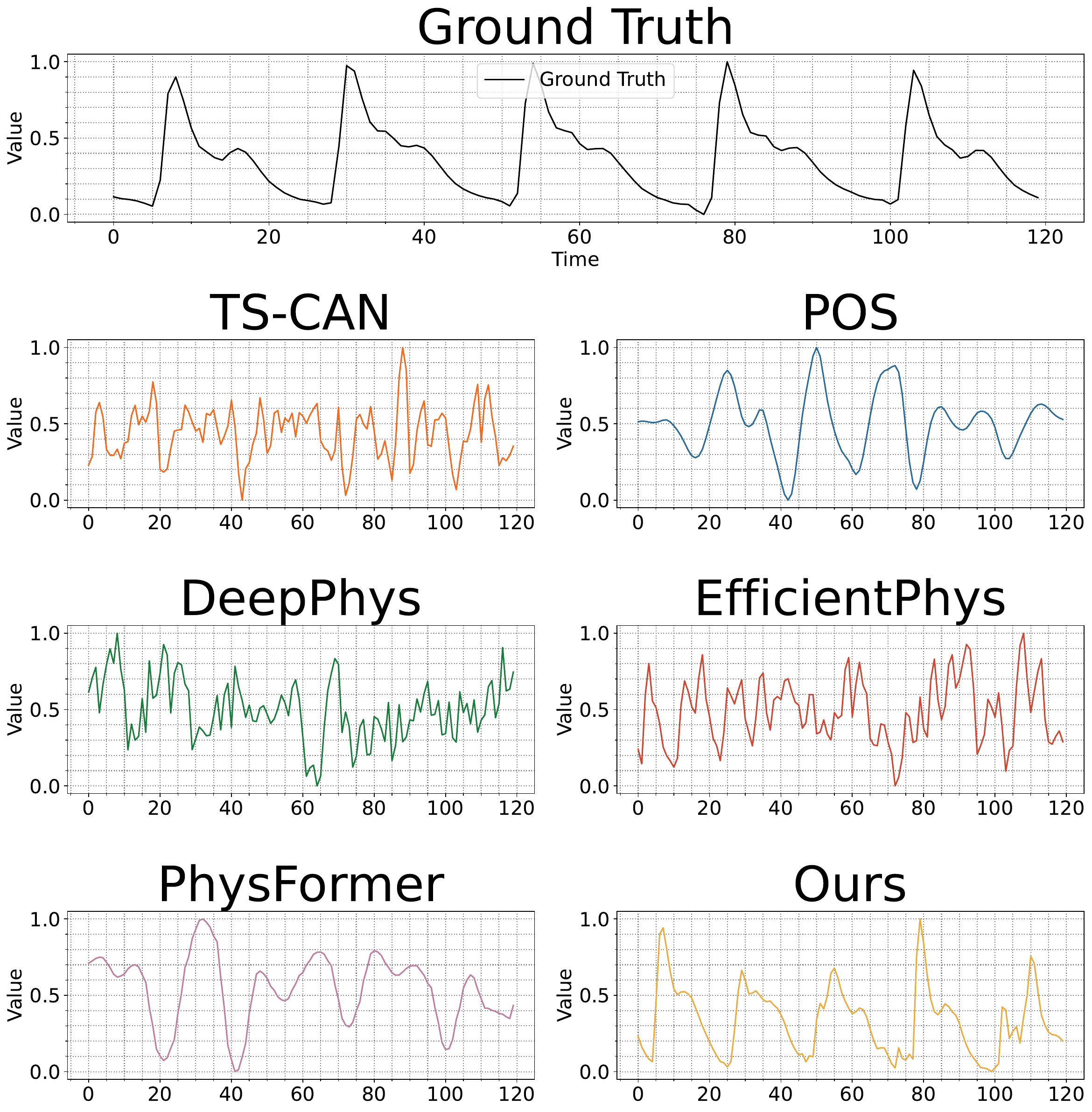}
        \caption{Talking}
        \label{subfig:c}
    \end{subfigure}\hfill
    \label{fig:mmpd_motion}
\end{subfigure}\hfill

\begin{subfigure}{\linewidth}
    \centering
    \begin{subfigure}{.32\linewidth}
            \includegraphics[width=\linewidth]{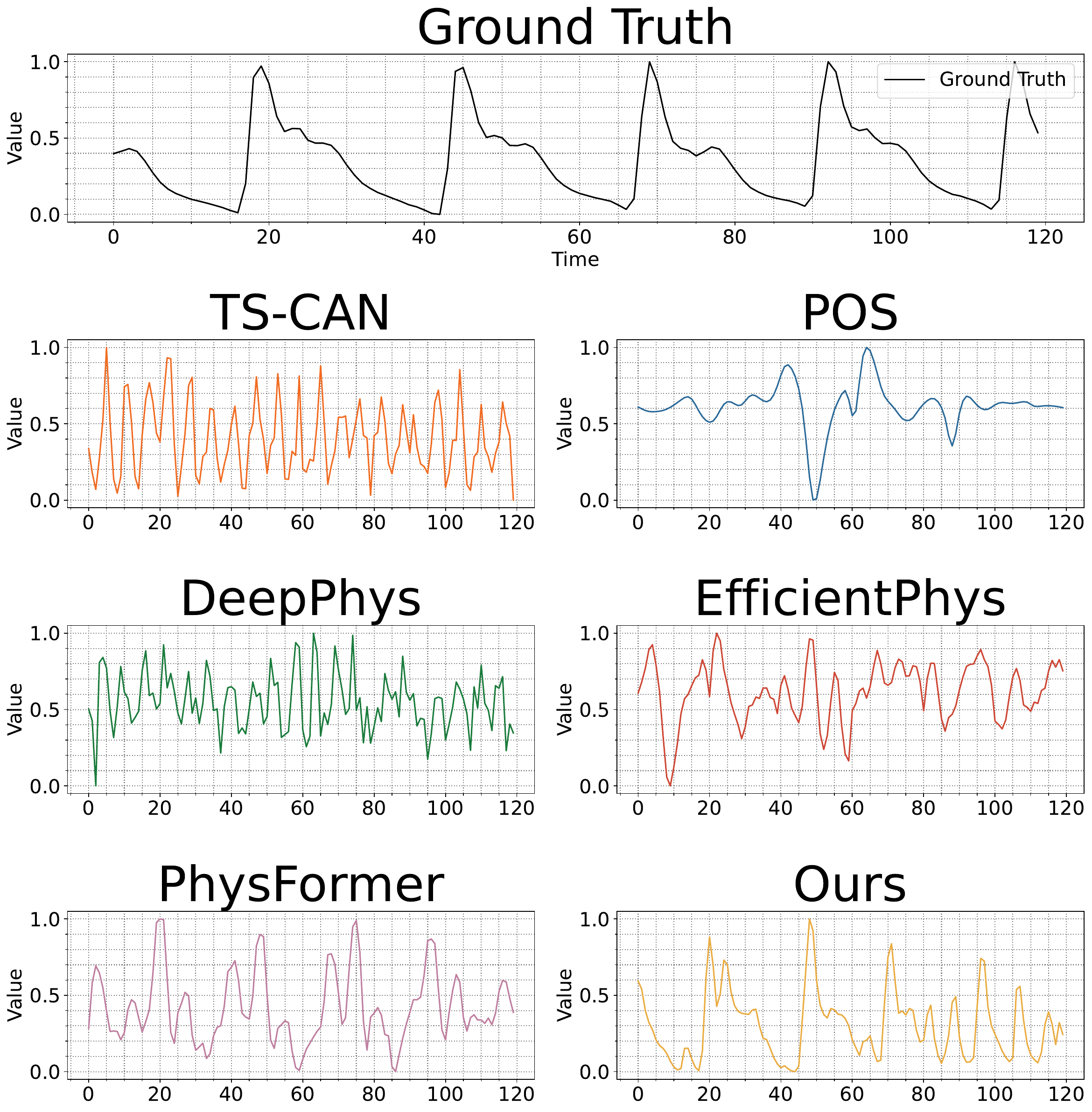}
            \caption{Low Light}
            \label{subfig:d}
    \end{subfigure}\hfill
    \begin{subfigure}{.32\linewidth}
        \includegraphics[width=\linewidth]{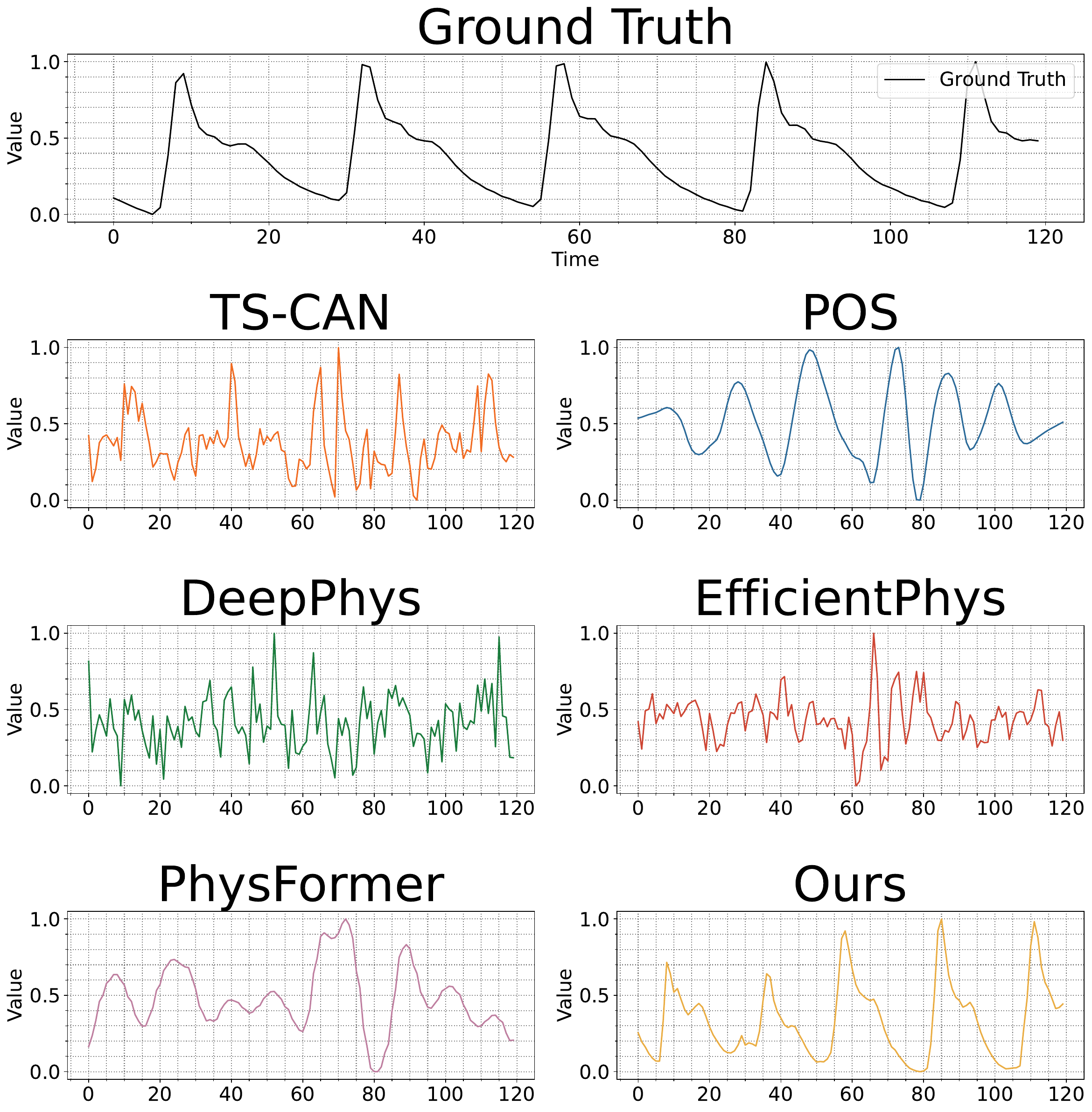}
        \caption{Incandescence}
        \label{subfig:e}
    \end{subfigure}\hfill
    \begin{subfigure}{.32\linewidth}
        \includegraphics[width=\linewidth]{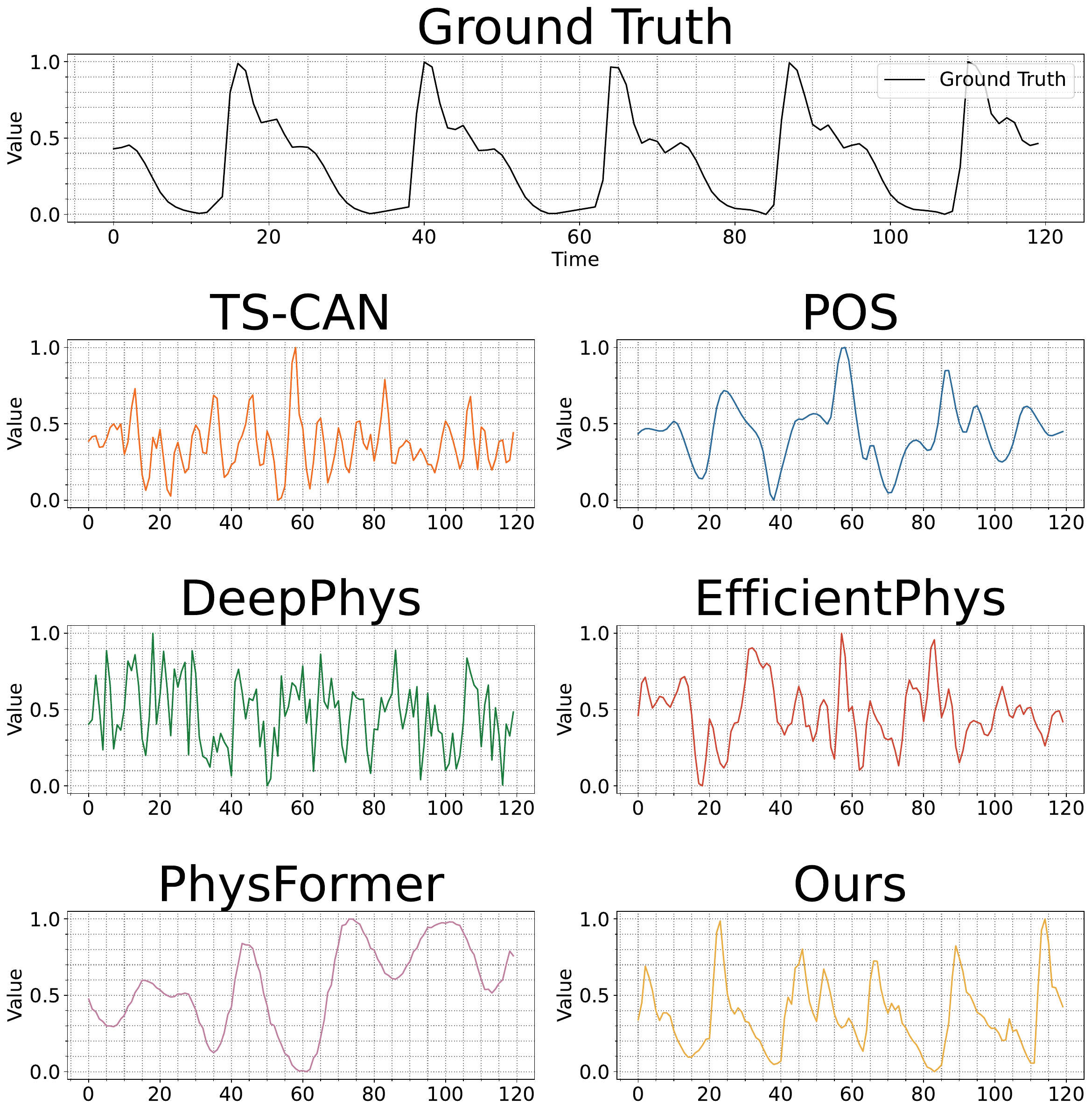}
        \caption{Nature Light}
        \label{subfig:f}
    \end{subfigure}
    \label{fig:mmpd_light}
\end{subfigure}
\caption{The performances of video-based approaches vary under different body movements and light conditions.}
\label{fig: mmpd}
\end{figure*}

\section{Model Design}
\label{sec: cardionet}

\subsection{Kinetics for Cardiac Learning}
\label{subsec: principle}
\head{Principles} 
\tmc{
In this section, we introduce the principles of extracting cardiac information from video and audio data, as shown in \fig\ref{fig:kinetics}.  
The fundamental concept revolves around the variations in blood pressure caused by cardiac activities, which manifest as quasi-periodic deformations of blood vessels. 
Since blood vessels circulate blood throughout the body, including the face, lungs, and throat,  we can infer heart activity in these areas through video and audio analysis. 
Specifically, when a light illuminates the skin, subtle color variations caused by pulse-induced blood flow can be captured through video streams. Additionally, as the lungs supply airflow for vocal fold vibrations and the throat modulates voice production, subtle cardiovascular motions associated with these processes can be detected in human speech.

In video streams, when light hits the skin, subtle color changes from pulse-induced blood flow can be captured, as described by the Dichromatic Reflection Model (DRM) \cite{shafer1985using}. We define the Domain of Interest (DOI) of the facial areas as $\Pi \in \mathbb{R}^{N_v \times C \times H_f \times W_f}$, and $\Pi_{i,j} \in \Pi$ denotes the RGB pixels at the $i$-th row and the $j$-th column. To bridge the color with RGB values, we model the spectral relationship as:
\begin{equation}
\label{eq: video_principle}
    \Psi_{\Pi_{i,j}}(f) = I(f) \ast \Delta(f),
\end{equation}
where $I(f)$ is the illumination spectral components, $\ast$ is the convolution operation, and $\Delta(f)$ is the reflection modulator, comprising specular reflection $\Delta_s(f)$ and diffuse reflection $\Delta_d(f)$. Specular reflection occurs at the epidermis level, while diffuse reflection penetrates into the hypodermis, reflecting off capillaries and blood vessels, encapsulating the physiological spectrum $H(f)$. We further decompose $I(f)$ and $\Delta_s(f)$ into static and dynamic components, where dynamic components are denoted as $\mu(H(f), O(f))$ and $\nu(H(f), O(f))$, respectively. $O(f)$ is a set of irrelevant signals. $\mu(\cdot)$ and $\nu(\cdot)$ are transfer functions without analytic expressions. Our goal is to infer $h(t)$ from $\Pi$, where $h(t)$ is the temporal counterpart of the spectral representation $H(f)$.

Speech is a complex auditory phenomenon that carries biological information. The airflow is produced from the lungs, which is then modulated by the vocal folds within the larynx to generate sound. This sound is further shaped by the movements and positions of the articulatory organs, such as the tongue and throat. Formally, the speech signal $\Xi$ can be formulated in the frequency domain as 
\begin{equation}
\label{eq: audio_principle}
    \Psi_{\Xi}(f) = L(f) \cdot R(f),
\end{equation}
where $L(f)$ is the sound energy source. $R(f)$ is an acoustic filter creating formant, affected by the vocal tract's physical attributes. Blood flow in surrounding vessels, particularly carotid arteries, influences the acoustic properties \cite{xu2022hearing}. These cardiovascular dynamics are encapsulated in the model by integrating the physiological signal $\hat{H}(f)$ into $R(f)$. 
}

\head{Observations}
Existing video-based solutions \cite{liu2020multi, chen2018deepphys, yu2019remote, liu2023efficientphys, zou2024rhythmformer, wang2016algorithmic}, though many, are trained on small datasets with controlled environments, \eg, PURE \cite{stricker2014non}. Their performances will degrade greatly when training and testing on more complicated datasets, \eg, MMPD \cite{10340857}. As can be seen from  \fig\ref{fig: mmpd}, the existing video-based solutions cannot effectively capture the cardiac semantics across different body movements and light conditions. These results present a grand challenge for cardiac learning. Meanwhile, different light conditions and body movements will degrade the performance from the video-based approaches, where audio can help \cite{xu2022hearing}. Therefore, our goal is to design a dedicated audio-visual network to extract those motions.

\begin{figure}[t]
    \centering
    \includegraphics[width=\linewidth]{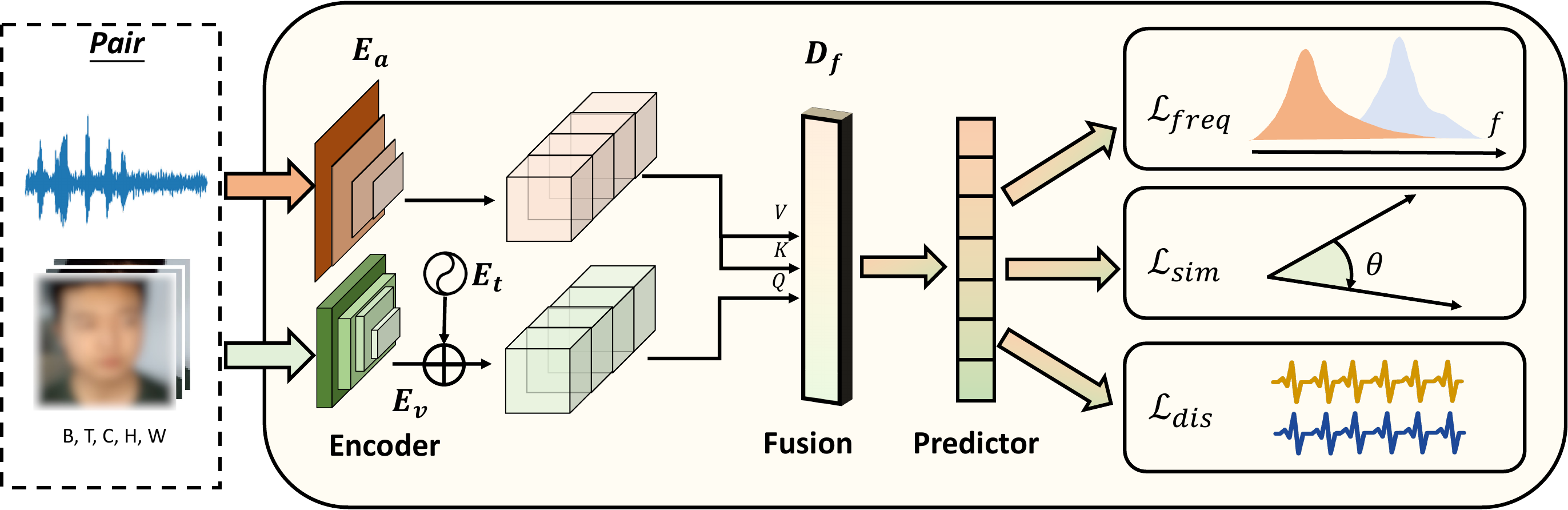}
    \caption{Overall Illustration of \texttt{CardioNet}}
    \label{fig:overall_cardionet}
\end{figure}

\subsection{\texttt{CardioNet} Design}
As shown in \fig\ref{fig:overall_cardionet}, the DOI pairs, \ie, frames $\Pi$ and audio clips $\Xi$, will be fed into video encoder $E_v$ and audio encoder $E_a$, respectively, followed by a fusion network to aggregate the two modalities.

\subsubsection{Video Branch Design}
We will first introduce $E_v$.

\head{Temporal Differential Block (TDB)} The input video frames $\Pi$  will first be processed as, \ie,
\(
    \dot{\Pi}_{i,j}^{t}=\Pi_{i,j}^t - \Pi_{i,j}^{t-1}.
\)
Since we only have past information, we perform backward differentiation. 
The key idea is, we treat the psychological activities as tiny local "motions". 
It efficiently captures the changes between consecutive frames \cite{wang2021tdn}. Furthermore, TDB plays a crucial role in isolating dynamic features while suppressing static components present in the video data.
Temporal difference enhances the contrast of the cardiac signal $h(t)$ within the latent space, facilitating more effective feature extraction and subsequent analysis. 
Thereafter, they are fed into convolution networks and upsampled to meet the length of video features. It is also imperative to capture the static information inherent in the video frames. To this end, we integrate a parallel pathway to process the original video frames, allowing for a more comprehensive understanding of the environment. We then introduce lateral connections to facilitate fusion of static and dynamic information.

\begin{figure}[t]
    \centering
    \begin{minipage}{0.48\linewidth}
        \centering
        \includegraphics[width=\linewidth]{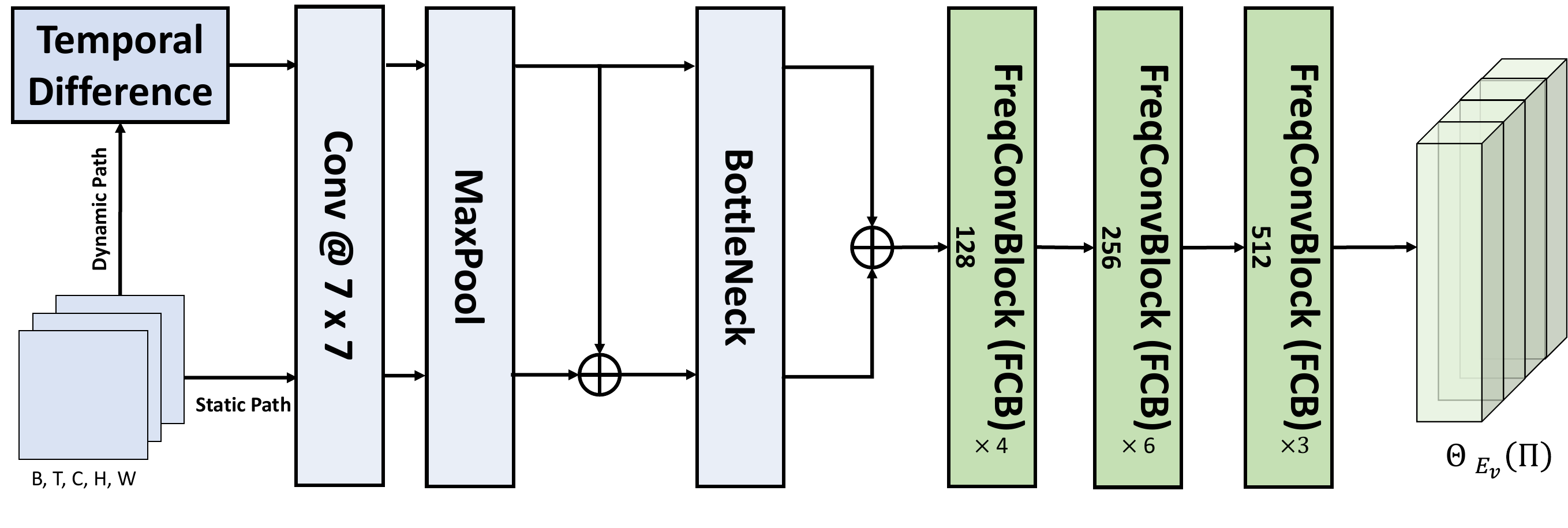}
        \caption{Video Encoder}
        \label{fig:video_enc}
    \end{minipage}\hfill
    \begin{minipage}{0.48\linewidth}
        \centering
        \includegraphics[width=\linewidth]{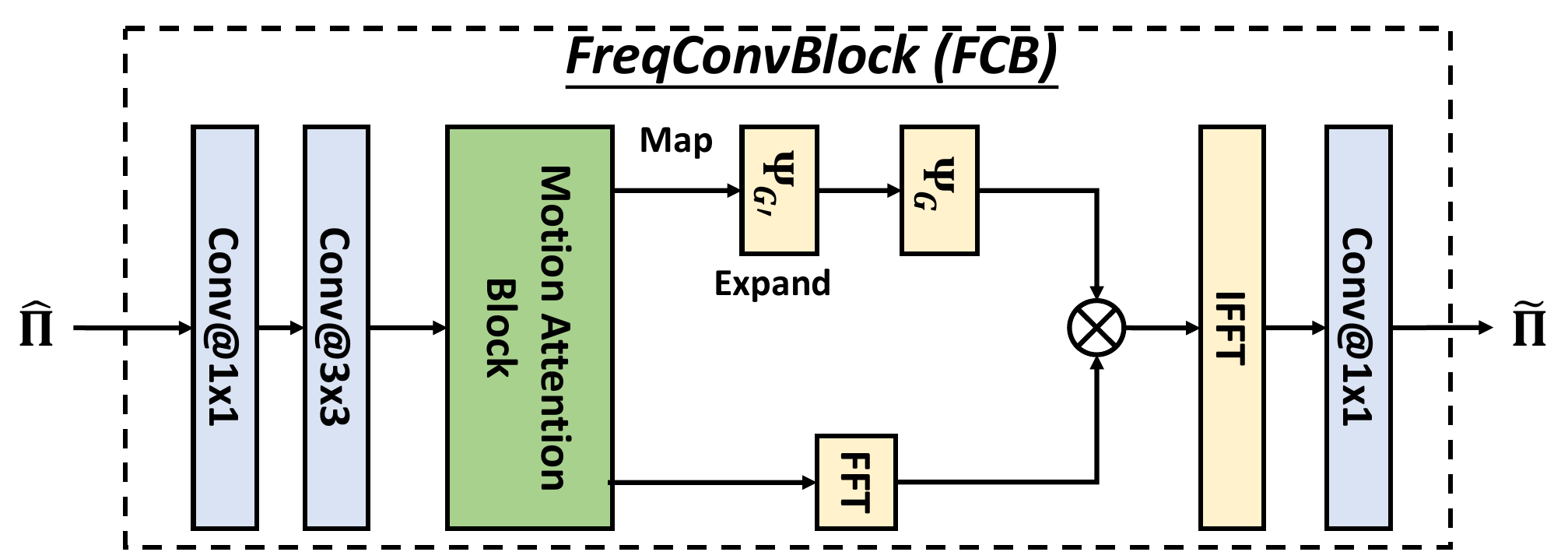}
        \caption{Frequency-Aware Convolution Block (FCB)}
        \label{fig:fcb}
    \end{minipage}\hfill
\end{figure}

\head{Motion-Aware Aggregation (MAA)}
After lateral fusion, we pass the intermediate latent to the bottleneck block to extract the spatial information and increase the expressive power. 
We recognize the importance of spatial modeling in mitigating the motion noise from head movement. 
Unlike video recognition tasks, where the relative location of the pixel is vital, we care more about how to track the variations of these pixels over time. 
To this end, we introduce a self-attention mechanism for frame-wise aggregation between consecutive frames. 
Our goal is to establish a mapping between temporal pixel variations and consecutive spatial information. Given the latent space $\boldmath{\hat{\Pi}}$, we query the one pixel at time $t$, \ie,  $\hat{\Pi}^{t}_{i,j}$ and compute the attention with previous frame,
\begin{equation}
    \rho^{t} = \operatorname{Softmax}\left(\frac{\hat{\Pi}^{t}_{i,j} \cdot \left(\hat{\Pi}^{t-1}_{i \pm \Delta i, j \pm \Delta j}\right)^T}{\sqrt{d_k}}\right).
\end{equation}
Here $\Delta i=\Delta j=k/2$, which is the perception grid size. $d_k$ is the dimension of $\hat{\Pi}^{t-1}_{i \pm \Delta i, j \pm \Delta j}$. $\rho^t$ captures the inter-frame pixel displacement, drawing attention to motion while enhancing temporal features between frames. Subsequently, we can get the weighted sum of temporal neighbor frames and aggregate with a query to enhance the original pixel:
\begin{equation}
    \check{\Pi}^{t}_{i,j} = \hat{\Pi}^{t}_{i,j} + \rho^{t} \cdot \hat{\Pi}^{t-1}_{i \pm \Delta i, j \pm \Delta j}.
\end{equation}
This mechanism scrutinizes pixel displacements across consecutive frames, akin to tracing the path of movement. Each pixel's attention weight encapsulates its significance in depicting motion, allowing the model to recognize subtle shifts and fluctuations over time.

\head{Frequency-Aware Block (FAB)}
After applying motion attention aggregation, we acquire the enhanced feature $\boldmath{\check{\Pi}}$. Our previous focus has been on modeling video dynamics in the temporal domain.
These are very effective designs for cardiac time series learning. Moreover, given the intrinsic property of $h(t)$, which turns out to be a quasi-periodic signal,  it becomes imperative to incorporate frequency features into our analysis. Here, the term "frequency" does not merely refer to the spectrum of color space within the video; rather, we aim to capture the underlying frequency variations of pixels over time. 
Inspired by DTF \cite{long2022dynamic}, we 
attempt to explicitly incorporate FFT in our design. For each pixel $\boldmath{\check{\Pi}_{i,j}} \in \mathbb{R}^{T \times \hat{C}}$, we apply FFT along the temporal dimension to acquire the feature spectrum $\Psi_{\boldmath{\check{\Pi}_{i,j}}}(f)$. 
To capture the frequency information, we introduce a learnable frequency filter $\Psi_{G}(f) \in \mathbb{R}^{\hat{C} \times N_{f}}$. We use IFFT to get the modulated temporal feature.
With FCB, we can enlarge the receptive field and profile cardiac time series with frequency constraints.

\head{Irregular Sampled Time Embedding} 
Another challenge is the fluctuating FPS. 
To this end, we add the timestamp feature to handle the irregular sampled time. Given the set of timestamps $\{t_i\}_{i=1}^{N_v}$, we design the timestamp embedding $E_t$ and fuse it with $\boldmath{\Theta_{E_v}(\Pi)}$. Specifically, we employ a frequency embedding scheme, which computes triangle embedding based on a geometric progression of frequencies up to $f_m$. We first derive a set of frequencies with the size of embedding dimension $N_{t_d}$, \ie,
\begin{equation}
    \omega^{k} = \exp \left(\frac{2k}{N_{t_d}}\cdot \log (f_m)\right),
\end{equation}
where $k=1, \cdots N_{t_d}/2$. Then the angle for each timestamp $i$ is given by 
\(
    \theta_{i}^{k} = t_i \cdot \omega^{k} \cdot 2\pi.
\)
Finally, the timestamps are embedded through trigonometric positional encoding.

\subsubsection{Audio Branch Design}
We then introduce the audio encoder. 

\head{Raw Audio}  Traditional audio-based learning often leverages mel-spectrogram. However, this method may not be suitable for our task. Our predictions, $h(t)$, manifest as quasi-periodic signals, ideally shown as straight lines on a mel-spectrum. But because cardiac activities are variable, these lines will exhibit randomness on a temporal-frequency map. Also, the location of the "straight" line has physical meanings, rather than a simple pattern. Therefore, we resort to learning from the raw audio signals directly. 
The key insight is, the process of producing speed from our vocal organs is composed of several acoustic filters, as indicated in \S\ref{subsec: principle}.  We can simulate the effect of filters and incorporate them in our design.

\head{Temporal-Frequency Filter (TFF)} 
The cardiac effect on the speech can be seen as a match filter. To this end, we adopt the SincNet  \cite{ravanelli2018speaker}, which can be expressed as,
\begin{equation}
    r_i(t, \theta) = 2 f_{i,2}^{\theta} \operatorname{sinc} (2 \pi f_{i,2}^{\theta} \cdot t) - 2 f_{i,1}^{\theta} \operatorname{sinc} (2 \pi f_{i,1}^{\theta} \cdot t).
\end{equation}
$f_{i,2}^{\theta}$ and $f_{i,1}^{\theta}$ denotes the two cutoff frequencies. We can treat the two cutoff frequencies as learnable parameters. 
We then perform a convolution between $r_i(t)$ and the raw audio $\xi(t)$. 
They will be fed into 1D convolution blocks for feature extraction.

\subsubsection{Fusion Block Design}
We now present the design of the fusion network. We opt for the late-fusion scheme, as the relationships between audio and cardiac activity, as well as video and cardiac activity, are not initially apparent. \tmc{Moreover, late fusion provides architectural flexibility. In scenarios where audio is absent (such as user silence), our system gracefully defaults to video-only inference while maintaining consistency.} Within the fusion block, we aim to address two challenges: 1) aligning the audio and video features along the temporal domain, and 2) handling the sampling rate mismatch between the audio and video features. To do so, we propose a multi-head temporal attention fusion block. Subsequently, the fused feature will be passed through linear fully connected layers.
Technically, we exploit video features as the query, and audio features as the key and value
, \ie,
\begin{equation}
    \Theta_{f}(\Pi, \Xi) = \operatorname{Softmax}\left(\frac{\Theta_{E_v}(\Pi) \cdot \Theta^T_{E_a}(\Xi)}{\sqrt{d_{E_v}}} \right) \cdot \Theta_{E_a}(\Xi).
\end{equation}
The fused feature $\Theta_f(\Pi, \Xi)$ will be fed to the output layer.

\begin{figure}[t]
    \centering
    \includegraphics[width=0.8\linewidth]{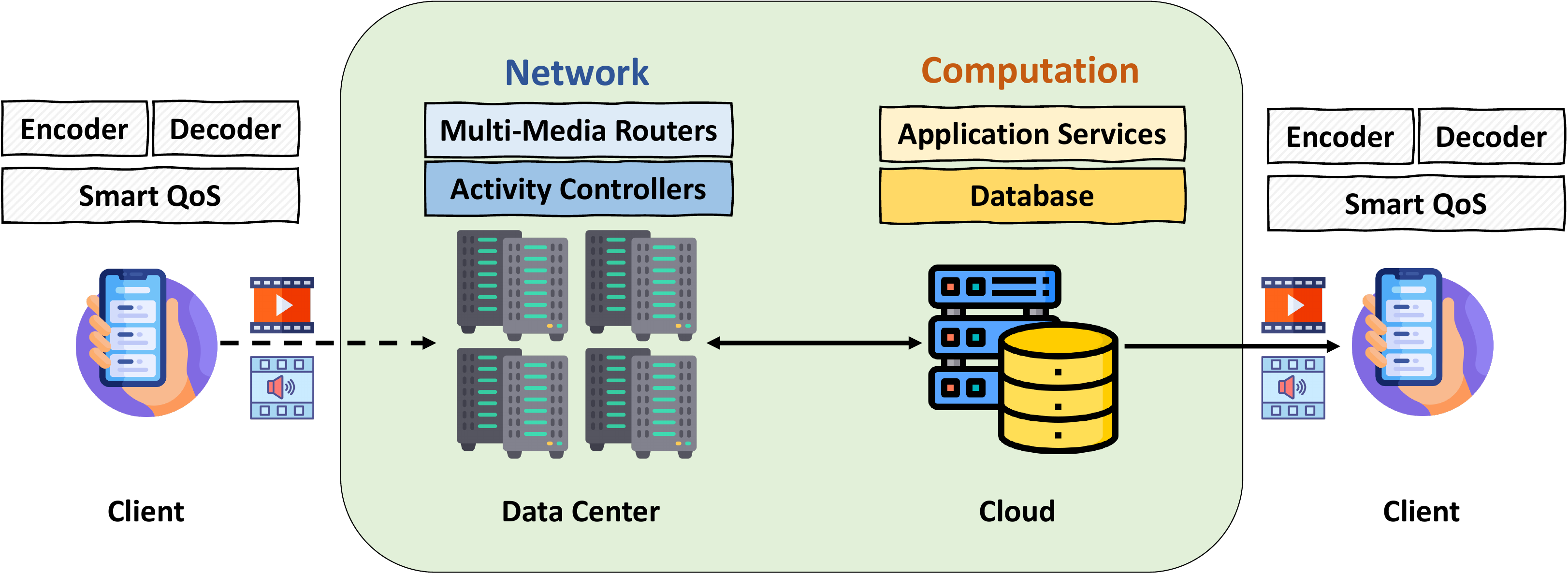}
    \caption{The architecture of a video streaming system.}
    \label{fig:streaming_arch}
\end{figure}
\subsubsection{Loss} 
In this part, we will elaborate our loss function design. 
We include three types of loss functions, \ie, focal loss, frequency loss and similarity loss,
\(
    \mathcal{L}_{\text{all}} = \alpha \cdot \mathcal{L}_{\text{dis}} + \beta \cdot \mathcal{L}_{\text{sim}} + \gamma \cdot \mathcal{L}_{\text{freq}},
\)
where $\alpha$, $\beta$ and $\gamma$ are weights to balance the loss items. The focal loss $\mathcal{L}_{\text{dis}}$ excels at keeping peaks in the physiological signals \cite{ross2017focal}. The similarity loss $\mathcal{L}_{\text{sim}}$ represents the extent of alignment. Additionally, as we are learning a quasi-periodic signal, we incorporate spectral loss $\mathcal{L}_{\text{freq}}$ as well by calculating the MSE of FFT.

\section{System Design}
\label{sec: sys_design}

\begin{figure}[t]
    \centering
    \begin{minipage}{.9\linewidth}
    \begin{subfigure}{.48\linewidth}
            \includegraphics[width=\linewidth]{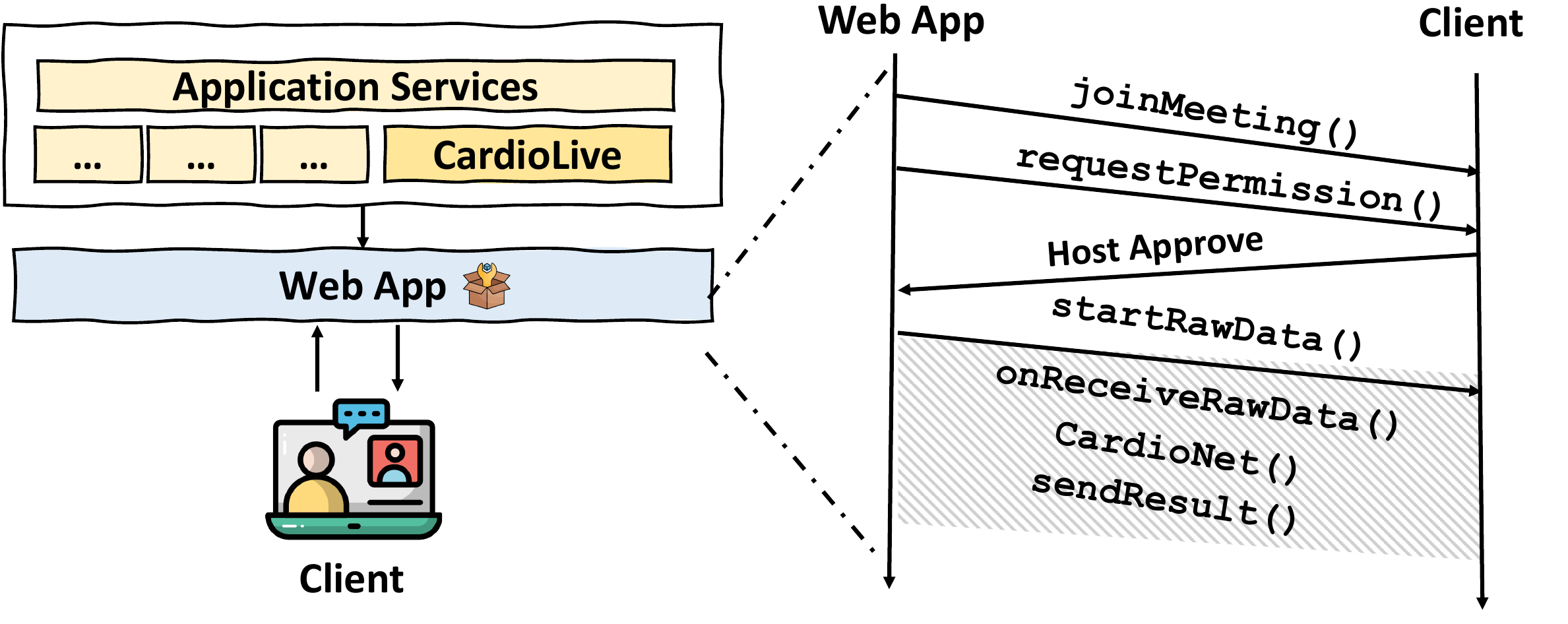}
        \caption{In-app bot}
        \label{fig:zoom_arch}
        \end{subfigure}\hfill
        \begin{subfigure}{.48\linewidth}
            \includegraphics[width=\linewidth]{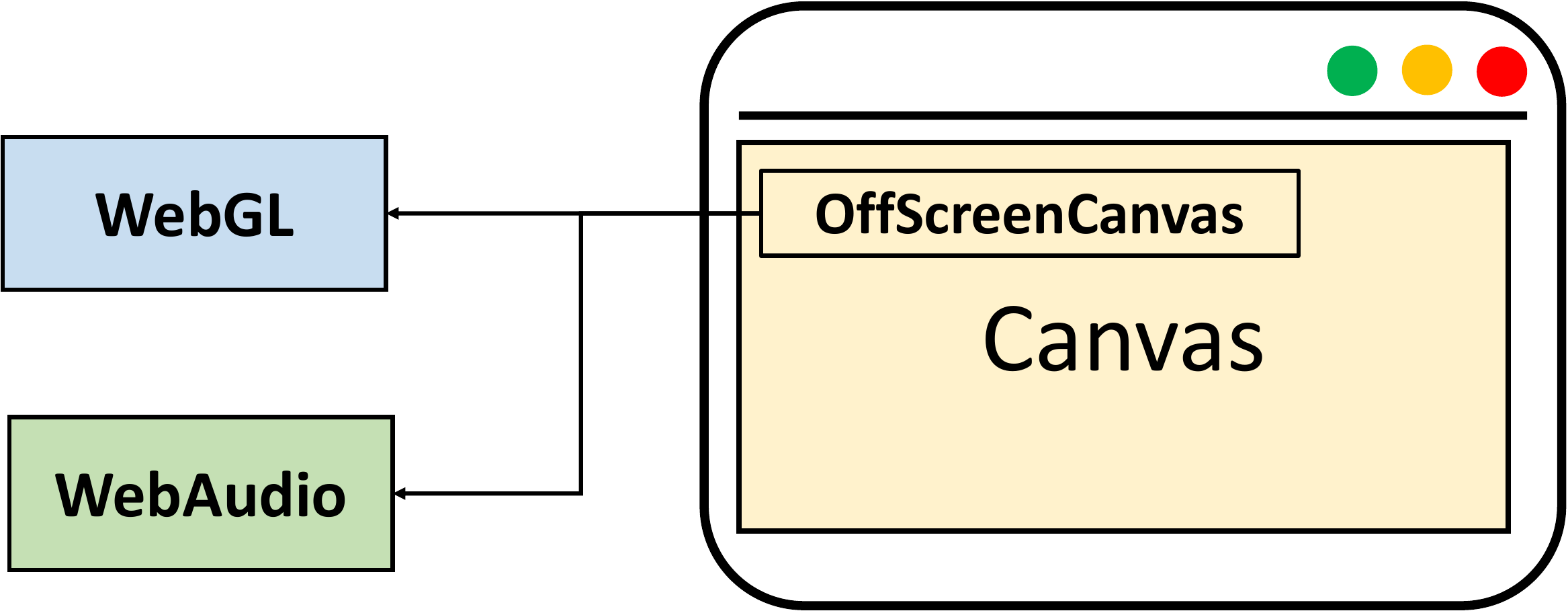} %
            \caption{Web extension}
            \label{subfig:chrome_av}
        \end{subfigure}
    \caption{Data Hook Design.}
    \label{fig:data_hook}
    \end{minipage}
\end{figure}
\begin{figure}[t]
    \centering
    \includegraphics[width=.9\linewidth]{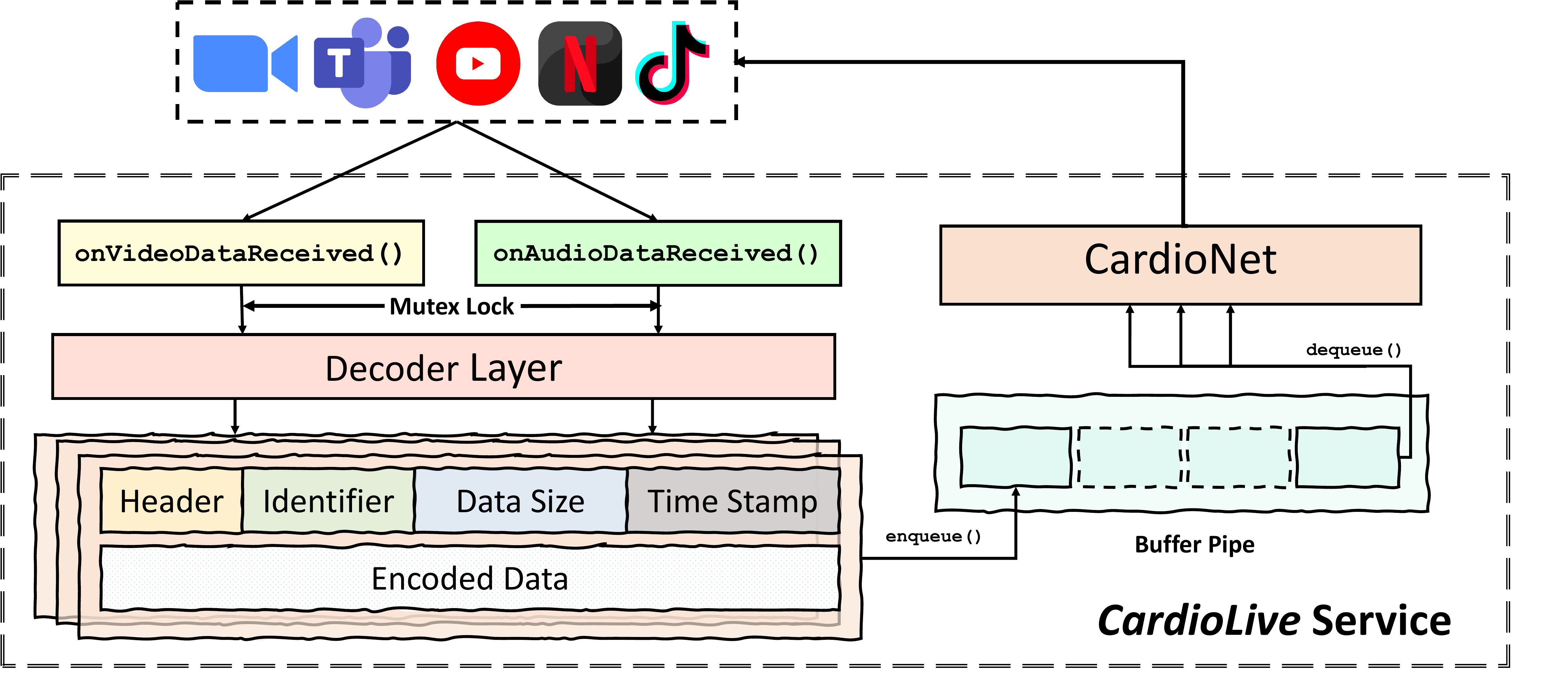}
    \caption{Packet and Buffer Design}
    \label{fig:packet_design}
\end{figure}

\subsection{Design Goal}
\label{subsec: design_goal}
Modern video streaming systems are complicated, and integrating OCM into them is non-trivial. 
As shown in \fig\ref{fig:streaming_arch}, the content is sent through cloud servers spanning across different locations globally. 
Besides running the data center and cloud computing, these video streaming systems offer a range of application services, such as content summarization, transcriptions, and AI-driven interactive features. %
For VoD providers, integrating new features is straightforward because they can preload resources in their data centers.  However, this does work well with streaming systems with live content interactions.
Meanwhile, deploying cardiac monitoring on end devices is also valuable. Users will be concerned about how the sensitive data are communicated over the network. 
To achieve SoD cardiac monitoring, we consider deploying it both on the ends and on the cloud. 
We package \sysname into a service, which both end users and manufacturers can readily access. 
In other word, we are not concerned about implementations on specific platforms, yet develop \sysname as a \textit{microservice}. We elaborate on it below.

\subsection{Buffer Design}
\label{subsec: buffer}
\head{Data Hook} We design data hooks to get the video and audio streams, namely \path{onVideoDataReceived()} and \path{onAudioDataReceived()}. Meeting platforms like Zoom usually support internal bots that join the calls. We can leverage the bots to access the raw data streams, as shown in \fig\ref{fig:zoom_arch}. Meanwhile, most of the video streaming systems are based on web pages, \eg, YouTube, Bilibili, \etc. Directly accessing the video streams of this platform is rather complicated and violates the policies. To this end, we leverage WebGL and WebAudio that exist in modern browsers to get the data streams, as shown in \fig\ref{subfig:chrome_av}. The browsers usually provide the Document Object Model (DOM), a programming interface to manipulate the structure, style, and content of web content. Our service will first access the canvas, an element for graphics on a web page, through the DOM. 
The canvas offers a bitmap where each pixel can be individually manipulated. We get the rendering context through WebGL and create an offscreen canvas that is rendered off the main thread and read the pixels through WebGL, preventing it from interfering with the normal UI updates. Meanwhile, we capture the audio from the video element through WebAudio, a versatile framework to handle audio operations on the web. We record the timestamp of the audio and video as well. Through the data hook, we can acquire the video and audio streams. Then we will construct them into data packets and buffer queues.

\head{Data Packet}
Normally, audio and video are encoded in separate ways. In meeting platforms, the video frames are usually encoded in YUV format, designed for the best transmission efficiency. 
To recover the original RGB streams and reduce the cost of decoding, we adapt a streaming-based decoding pipeline from GStreamer \cite{GStreamer}. We set the \verb|appsink| property for receiving the RGB data and assign \verb|appsrc| for handling YUV encoding. The transformation is in asynchronous mode so that the incoming frames will not conflict with the current operations. After that, we construct the collected frames in buffers. 
We feed the video-audio pair into the forwarding packets. For audio and video streams, we apply the same packet format, which contains a unique header, an identifier, the data size, timestamps, and the encoded payload data, as illustrated in \fig\ref{fig:packet_design}. The unique header is designed to judge whether the packet is correctly constructed and not mixed with other packets.
The identifier is assigned to indicate the audio or video data packets. We embed the received timestamps to denote the sequence of the video and audio, which will be further used for synchronization.

\begin{figure}[t]
    \centering

    \begin{minipage}{\linewidth}
        \begin{subfigure}{.48\linewidth}
            \includegraphics[width=\linewidth]{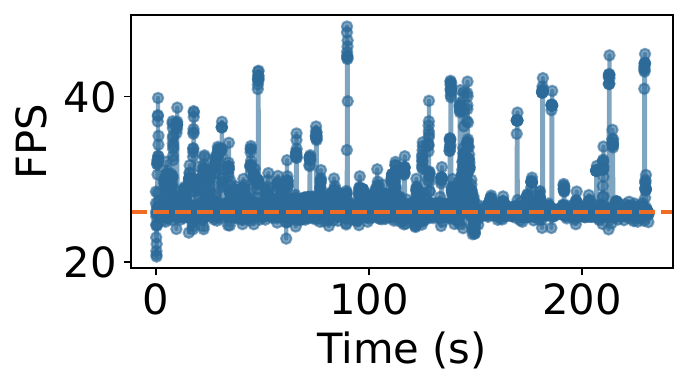}
            \caption{Chrome}
            \label{subfig:FPS_chrome}
        \end{subfigure}\hfill
        \begin{subfigure}{.48\linewidth}
            \includegraphics[width=\linewidth]{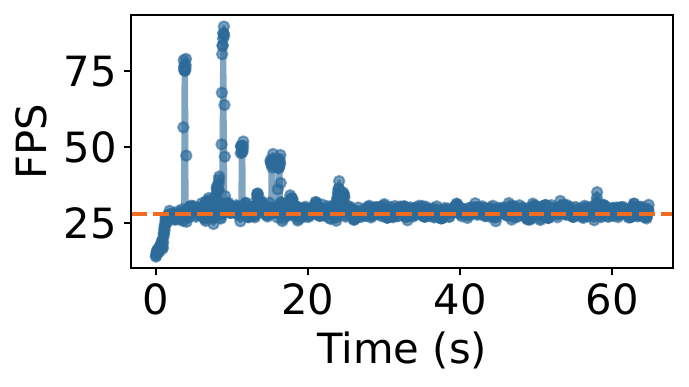} %
            \caption{Zoom}
            \label{subfig:FPS_zoom}
        \end{subfigure}
    \end{minipage}
    \caption{The FPS vary and change rapidly.}
    \label{fig: FPS}
\end{figure}
\begin{figure}[t]
    \centering
    \begin{minipage}{\linewidth}
        \begin{subfigure}{0.68\linewidth}
            \centering
            \includegraphics[width=\linewidth]{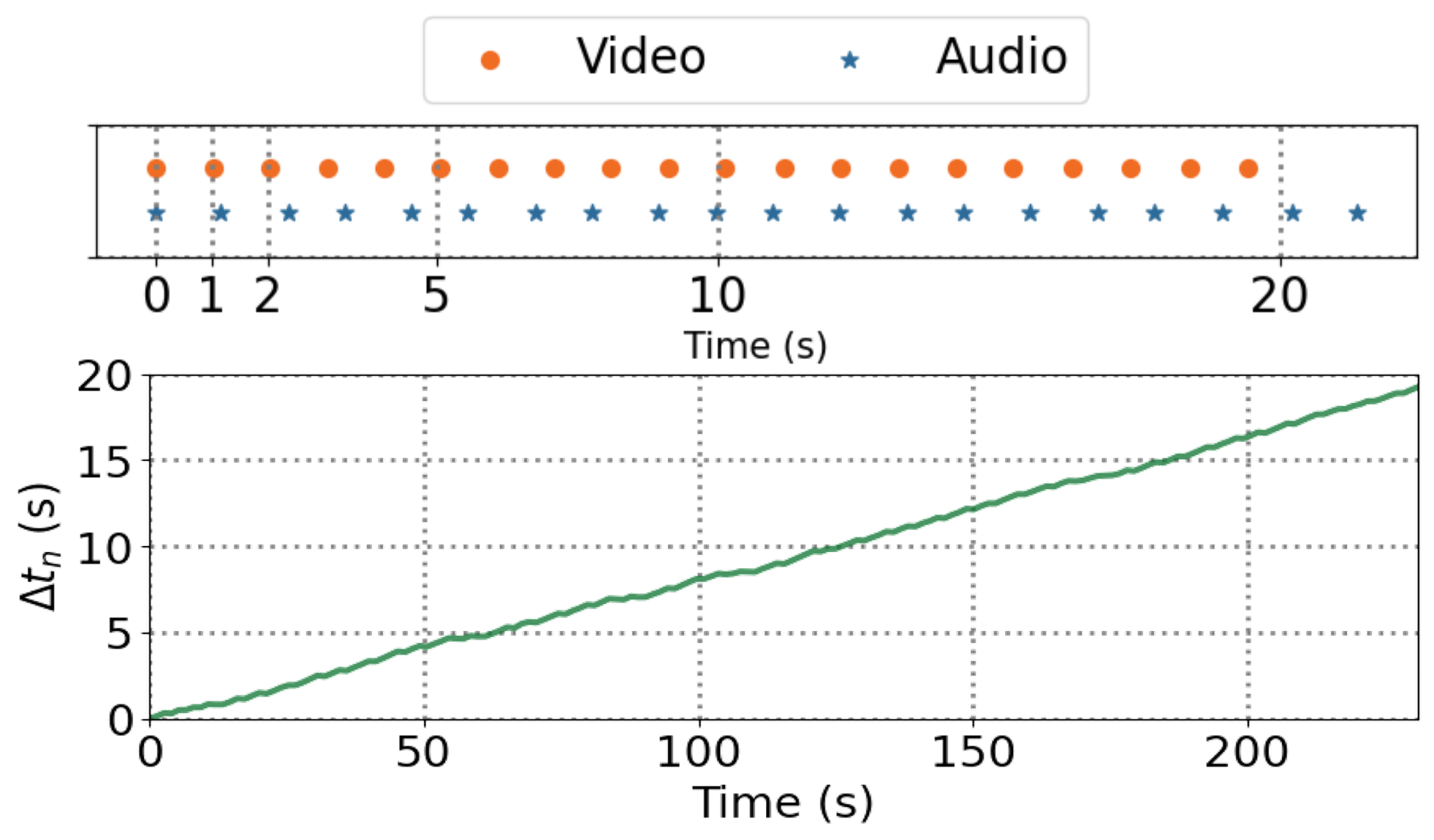}
            \caption{The temporal drifting}
            \label{fig:temporal_drifting}
        \end{subfigure}
        \begin{subfigure}{0.3\linewidth}
            \includegraphics[width=\linewidth]{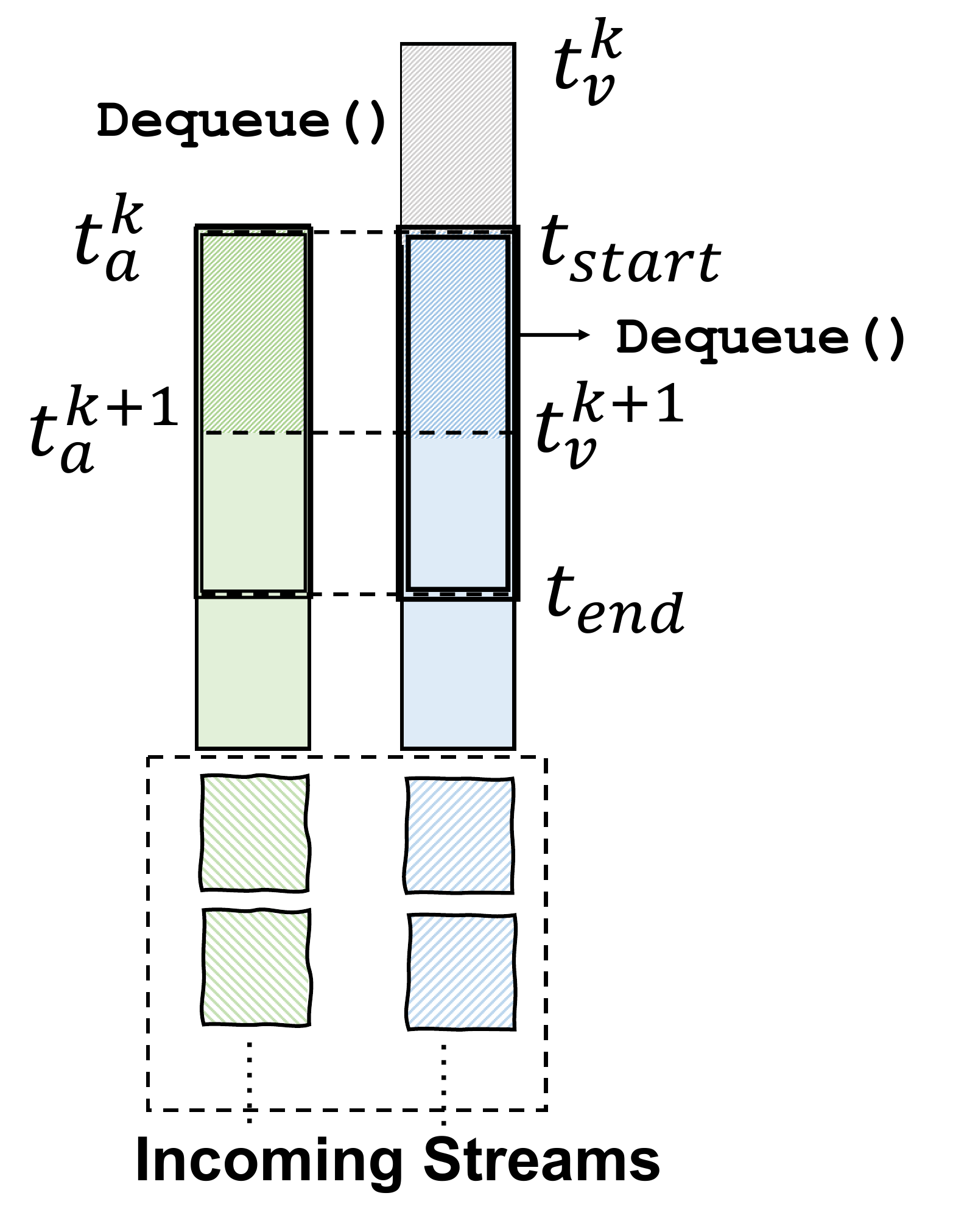} %
            \caption{Sync}
            \label{subfig:sync_scheme}
        \end{subfigure}
    \end{minipage}
    \caption{Audio-Video Synchronization Scheme.}
    \label{fig: av_sync}
\end{figure}
\begin{figure}[t]
    \centering
    \includegraphics[width=\linewidth]{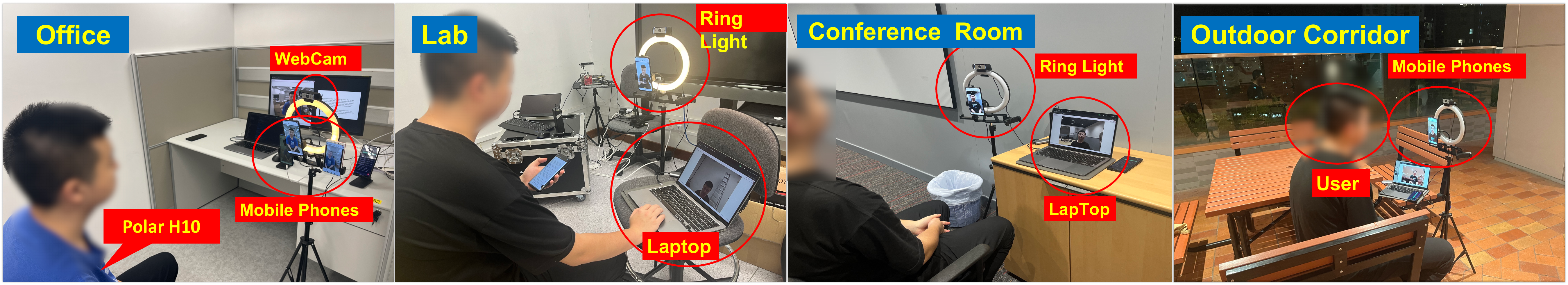}
    \caption{Experimental Setups}
    \label{fig:exp_set}
\end{figure}

\subsection{Service Design}
\label{subsec: service}
We abstract our system as a plug-and-play service. 
Our service first gets the hooked video and audio packets as the input. 
The data will be fed to the inference engine for output. 
We observe and tackle the two challenges: fluctuating FPS and unsynchronized streams.

\head{Drifting FPS} The fluctuating FPS will lead to two subproblems. 
Initially, the video streaming systems will ideally have 30 FPS but in reality undersampled at the receiver's end, as illustrated in \fig\ref{fig: FPS}, with some outliers present as well. 
Additionally, the frame rate is not constant, resulting in a varying number of frames within a given window. However, our model assumes a fixed 4-s input, with 120 frames of video (30 FPS) and 32000 samples of audio (8kHz). In other words, we have to adapt the real input size to the model.
To this end, instead of padding empty frames at the end, we duplicate a single frame circularly. For instance, if the actual FPS is 25, we insert an additional identical frame after every 5 frames to approximate a smoother transition to 30 FPS. Any remaining gaps at the end of the sequence are filled by repeating the last frame. As for overlarge FPS, we downsample the frames.  
For the audio clips, as 8kHz is much lower than the typical sampling rates (usually 32kHz or 44.1kHz) in modern video streaming systems, we can concatenate the received audio chunks and safely downsample them to 8kHz. 
\begin{figure*}[t]
    \centering

    \begin{minipage}{\linewidth}
        \centering
        \includegraphics[width=\linewidth]{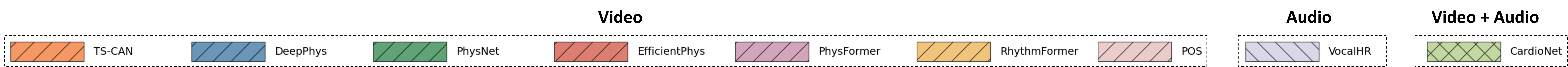}
    \end{minipage}

    \begin{minipage}{\linewidth}
        \begin{subfigure}{.32\linewidth}
            \includegraphics[width=\linewidth]{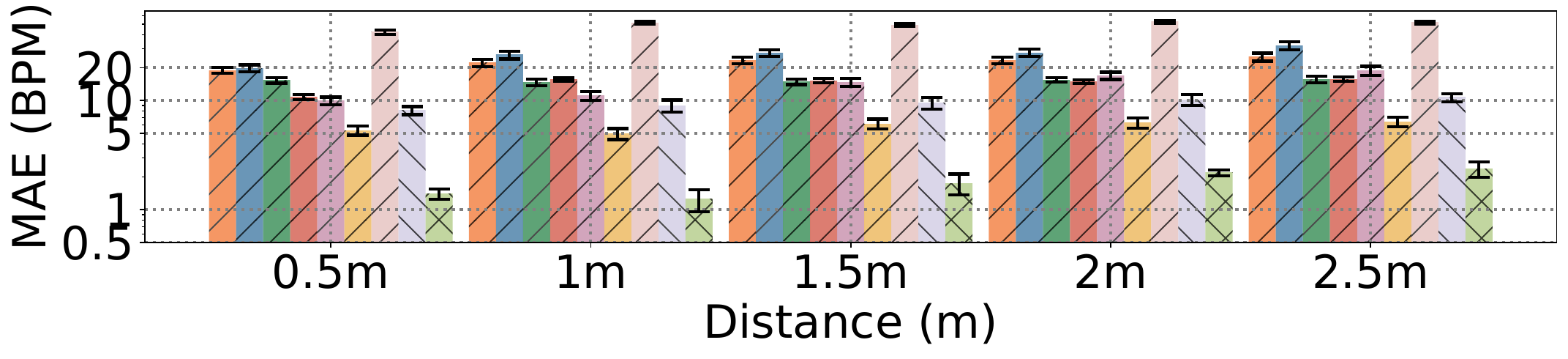}
            \caption{MAE}
            \label{subfig:dis_mae}
        \end{subfigure}\hfill
        \begin{subfigure}{.32\linewidth}
            \includegraphics[width=\linewidth]{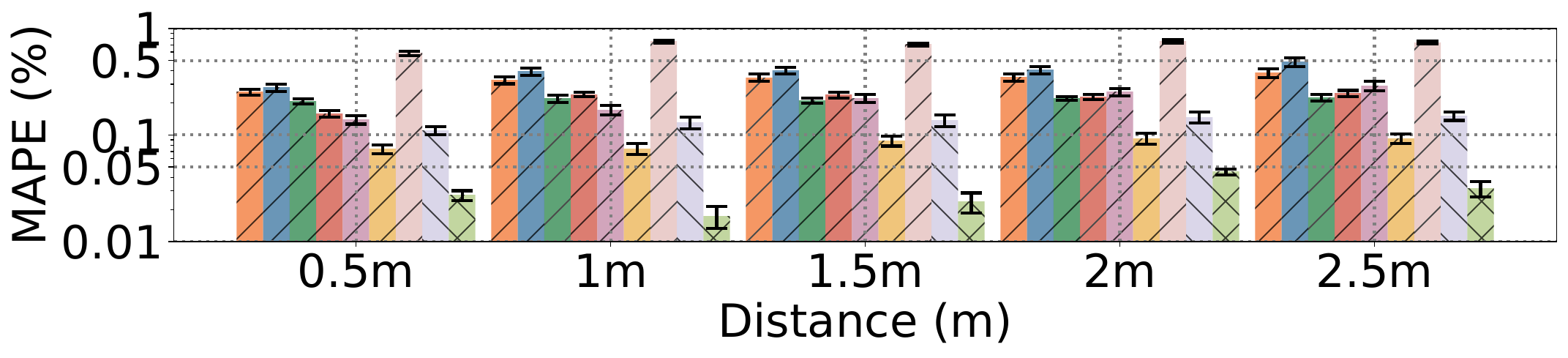}
            \caption{MAPE}
            \label{subfig:dis_mape}
        \end{subfigure}\hfill
        \begin{subfigure}{.32\linewidth}
            \includegraphics[width=\linewidth]{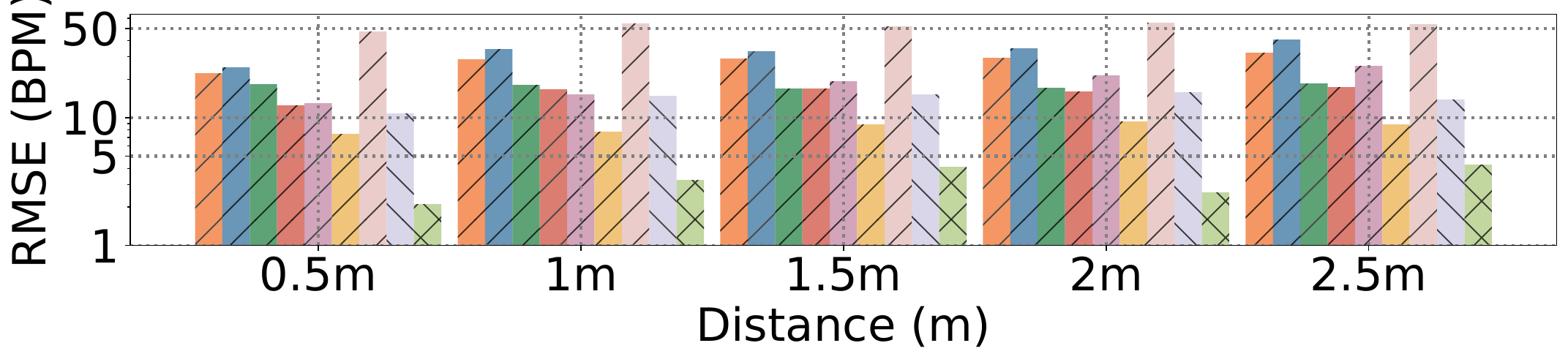}
            \caption{RMSE}
            \label{subfig:dis_rmse}
        \end{subfigure}
    \end{minipage}
    \caption{The performances for different distances.}
    \label{fig: res_dis}
\end{figure*}
\begin{figure*}[t]
    \centering
    \begin{minipage}{\linewidth}
        \begin{subfigure}{.32\linewidth}
            \includegraphics[width=\linewidth]{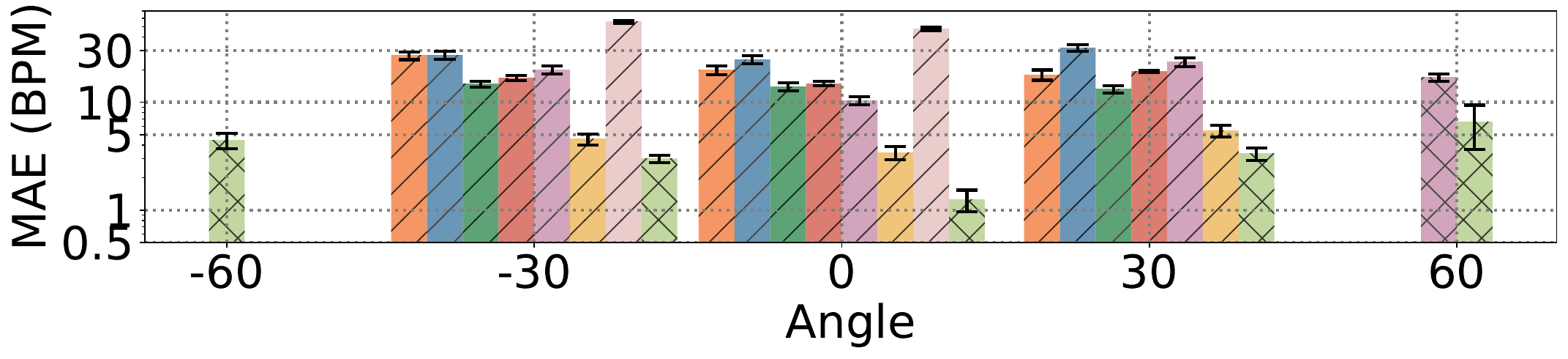}
            \caption{MAE}
            \label{subfig:angle_mae}
        \end{subfigure}\hfill
        \begin{subfigure}{.32\linewidth}
            \includegraphics[width=\linewidth]{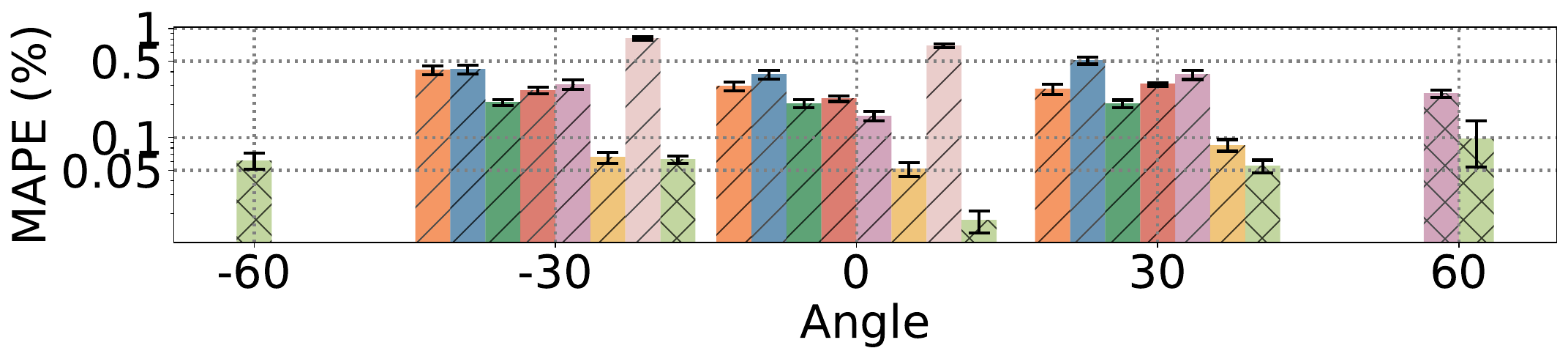}
            \caption{MAPE}
            \label{subfig:angle_mape}
        \end{subfigure}\hfill
        \begin{subfigure}{.32\linewidth}
            \includegraphics[width=\linewidth]{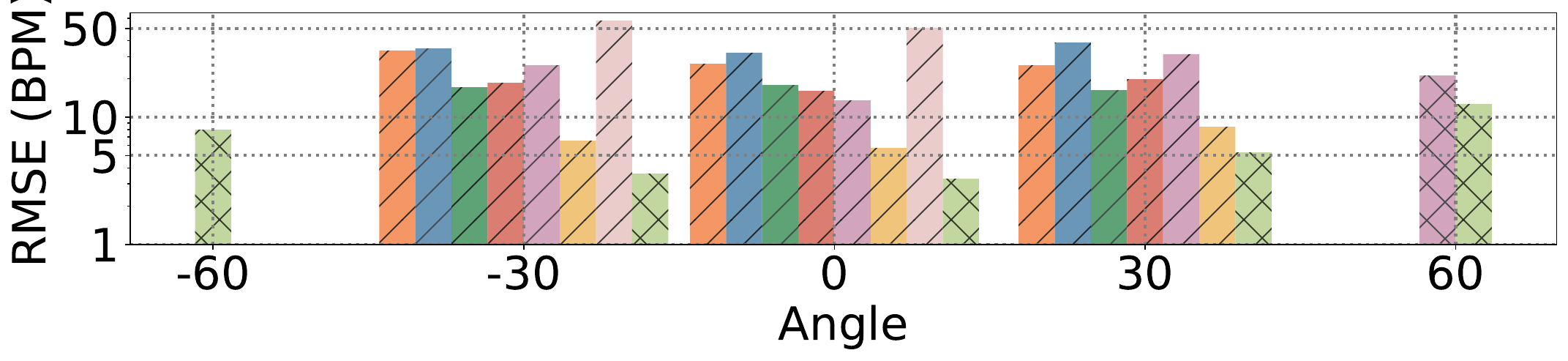}
            \caption{RMSE}
            \label{subfig:angle_rmse}
        \end{subfigure}
    \end{minipage}
    \caption{The performances for different angles.}
    \label{fig: res_angle}
\end{figure*}

\head{Audio Video Synchronization}
The audio-visual misalignment is a more severe issue. 
As the hooked audio and video are from separate channels, they will lose synchronization with the increase of time. As can be seen from \fig\ref{subfig:chrome_av}, the starting time of the audio and video will be misaligned quickly with accumulating drifts. To overcome this issue, we develop a scheme to ensure the audio and video chunks are synchronized before the inference engine. Given the audio and video streams $S_a(t)$ and $S_b(t)$, they will be extended to the buffer queues $Q_a(t)$ and $Q_v(t)$, respectively. We also maintain $t_a$ and $t_v$ as the starting time of audio and video chunks, respectively. We denote $\Delta t_n = t_a^{k} - t_v^{k}$ as the temporal drift between audio and video streams at the $k$-th trial. To mitigate the continuously increasing $\Delta t_n$, we align the start time at each step $k$ as, 
$t_{\text{start}}^{k} = \max(t_{a}^{k}, t_{v}^{k}),$
when $\Delta t_n$ is larger than the threshold $\epsilon_t$. We use $\epsilon_t=0.3$s. Then the ending time will be determined by $t_{\text{end}}^{k} = t_{\text{start}}^{k} + t_w$,
where $t_w$ is the window lengths. Note that we adopt a sliding window scheme, with window length $t_w$ and step length $t_s$. For the next window, the start time will be updated by finding the timestamp closest to, $t_{a}^{k+1} = t_{a}^{k} + t_s$ and $t_{v}^{k+1} = t_{v}^{k} + t_s$.
 Meanwhile, we will pop the items that have been processed from the buffer queues, \ie, 
 $Q_{a}(t) = Q_{a}(t) \textbackslash \{S_a(t) | t < t_{a}^{k+1}\}$ and $Q_{v}(t) = Q_{v}(t) \textbackslash \{S_v(t) | t < t_{v}^{k+1}\}$.
We then feed the synchronized pairs for inference.

\tmc{
\subsection{Preprocessing}
\label{subsec: preprocess}
In this section, we will discuss the preprocessing pipelines. 
We use the OpenCV face detector to find faces. We also perform voice activity detection to segment the talking period. Additionally, we need to separate multiple persons, if any, and match their audio and videos.

\head{Multi-person Separation}
We deduct the more challenging multi-user case into the single-user case by separating them. 
Initially, face detection can determine the number of participants. To ensure facial resolution, we focus on the largest $N_f$ faces, disregarding the others. 
Similarly, we will only consider $N_f$ speech clips with the largest power spectrum when separating audio. For efficiency, we choose $N_f=2$ in our paper. At this stage, the separated faces and speech segments may not correspond to each other. To address this mismatch, we proceed with audio-visual matching as described next.

\head{Audio-Visual Matching}
To realize the matching between speaking clips and facial hints, we adopt a cross-attention scheme \cite{tao2021someone, jiang2023target}. Specifically, after the encoders, we get two features $M_a$ and $M_v$. These features are expected to encapsulate relevant speaking activities by employing temporal encoders \cite{hu2018squeeze, afouras2018conversation}. To fuse the audio and video features, the audio features $M_a$ are integrated with the video data by treating $M_v$ as the target for querying through an attention framework. Conversely, the video features $M_v$ interact with $Q_a$, representing the audio query sequences. The outputs are concatenated together along the temporal direction.

}

\section{Evaluation}
\label{sec:evaluation}
In this section, we systematically evaluate \sysname. 
We perform comparison studies with the state-of-the-art (SOTA) video-based solutions and audio-based solutions. We mainly leverage our self-collected dataset. We use the following metrics to evaluate the accuracy of the model: Mean Absolute Error (MAE), Root Mean Square Error (RMSE), and Mean Absolute Percentage Error (MAPE).

\head{Data Collection} There is no existing dataset that can fit our requirements, with audio-visual pairs and clear heart rate ground truth. \tmc{Specifically, BP4D+ \cite{zhang2016multimodal} provides additional IR images but not the necessary audio; MMSE-HR offers only video; and while MAHNOB-HCI contains audio, it does not require participants to speak—only incidental utterances are present, which does not fit our scope.}
Therefore, we self-collect the dataset through 8 commodity mobile and laptop devices.
We leverage Polar H10 \cite{PolarH10Polar} to collect the ground truth. 
We recruit 10 users of diverse genders and skin colors\footnote{We have gained IRB from our university board.}. They are requested to read 10 materials \cite{sun2021ultrase}. Each round lasts for 40 minutes. 
We use a tripod along with a ring light to cast different light sources on the users. We collect overall of 84,666 data clips, which are clipped into facial regions with 4-s windows. We resize the video frames to 72$\times$72$\times$3 and the audio is resampled to 8kHz. The missing frames will be duplicated adopting the same scheme as \S\ref{subsec: service} mentions.  Moreover, we also make use of two publicly available video-only datasets: PURE \cite{stricker2014non} and MMPD \cite{10340857}.

\head{Software} We implement \texttt{CardioNet} through Pytorch. The model is trained via a single-card NVIDIA A100 80GB. We train the model with the learning rate of 1e-3,  AdamW optimizer, batch size of 16, and OneCycle scheduler. We use JIT to compile the model. We write 2000+ lines C++ code to implement the service in Zoom and 1500+ lines of JavaScript code for developing the service in the extension.

\head{Deployment}
We propose two deployment paradigms, web-based and app-based. For the web-based one, we develop a browser extension that operates \sysname in the background, which continuously captures audio and video data for processing, with results displayed on a canvas within the interface. In the app-based deployment, we register a bot \textit{in compliance with} the policies of the video streaming companies, which joins the sessions as a member, with the consent of all members. The data hook extracts audio and video for inference engines. The processed results are delivered through a notification system. Notably, the inference can be performed either on the company's cloud server or locally on the user's device.  In our real-world evaluation, we perform inference on the end device to show the robustness and efficiency.
\tmc{A demo video of \sysname can be found here: https://youtu.be/xoLmxPD264g.}

\subsection{Comparative Study}
We compare our \texttt{CardioNet} with various baselines. We choose the SOTA video-only baselines: TS-CAN \cite{liu2020multi}, DeepPhys \cite{chen2018deepphys}, PhysNet \cite{yu2019remote}, EfficientPhys \cite{liu2023efficientphys}, RhythmFormer \cite{zou2024rhythmformer}, POS \cite{wang2016algorithmic}. The last one is the signal processing method. We also reimplement VocalHR \cite{xu2022hearing}, the recent work that employs human speech for detecting heart rate. Through this study, we will justify our superior performances using both audio and video modalities.

\head{Distances} We first experiment with different distances from 0.5m to 2.5m. We apply the log-10 scale to each graph. 
As shown in \fig\ref{fig: res_dis},
while the error increases with distance for all methods, our approach consistently outperforms other baseline models at all tested distances. \texttt{CardioNet} achieves a MAE of just 1.40 BPM at 0.5m, significantly lower than the SOTA video-based baseline, \ie, RhythmFormer, by 73.7 \%, and 96.7\% lower than the worst-performing model, \ie, POS. Meanwhile, the audio-based model VocalHR has an MAE of 8.12 BPM at the same distance, which is 82.8\% higher than ours. 
Even at the maximum testing distance of 2.5 meters, \texttt{CardioNet} is still 63.1\% better than RhythmFormer and 77.9\% better than VocalHR. This demonstrates that the fusion of audio and video signals in \texttt{CardioNet} significantly enhances the overall performance. Besides, we observe the identical patterns of MAE, MAPE and RMSE, we will mainly report MAE for simplicity.

\head{Angles}
We evaluate our model across a range of angles from 0° to $\pm$60° at 1 meter, as shown in \fig\ref{fig: res_angle}. As it increases, video-based methods suffer from significant performance degradation due to reduced visibility of facial features. However, \texttt{CardioNet}, through audio-visual fusion, maintains robust performance across all angles. While the video quality deteriorates with extreme angles, audio signals remain unaffected by viewing angles. Even at $\pm$60°, where video signals typically falter, our model achieves up to 38.9\% lower MAE compared to baseline models. This result underscores the critical role of the audio modality at extreme angles.

\begin{figure}[t]
    \centering
    \begin{minipage}{\linewidth}
        \begin{subfigure}{.19\linewidth}
            \includegraphics[width=\linewidth]{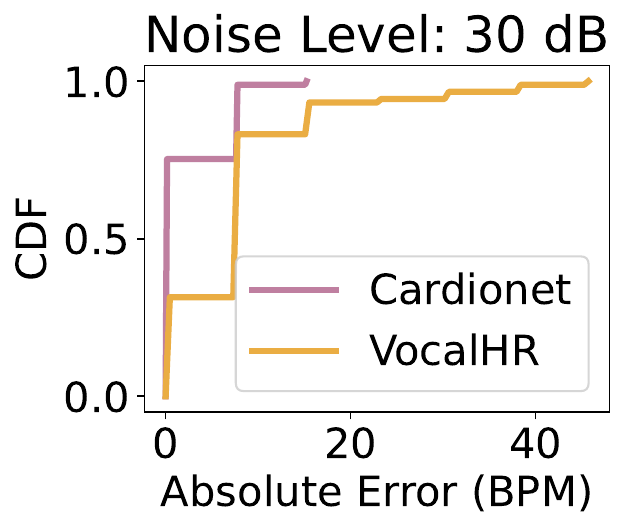}
            \label{subfig:noise_30}
        \end{subfigure}\hfill
        \begin{subfigure}{.19\linewidth}
            \includegraphics[width=\linewidth]{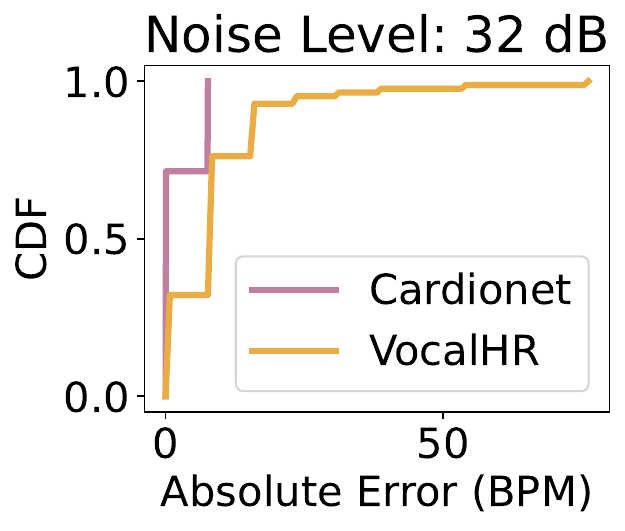}
            \label{subfig:noise_32}
        \end{subfigure}\hfill
        \begin{subfigure}{.19\linewidth}
            \includegraphics[width=\linewidth]{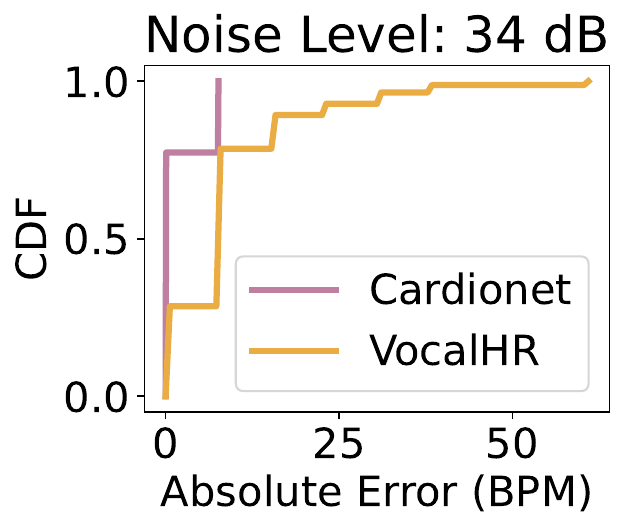}
            \label{subfig:noise_34}
        \end{subfigure}\hfill
        \begin{subfigure}{.19\linewidth}
            \includegraphics[width=\linewidth]{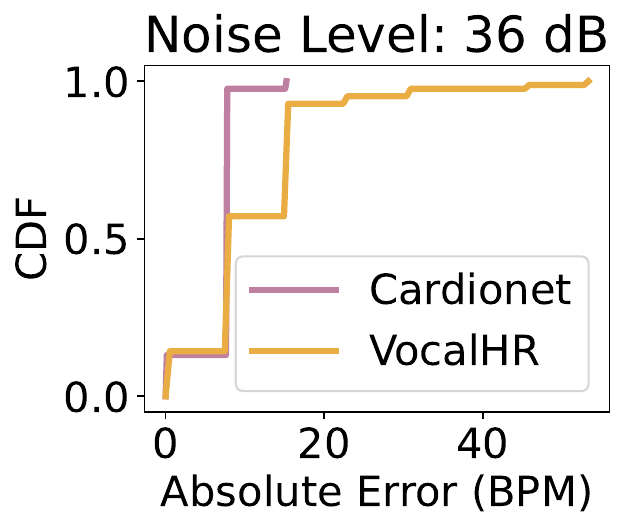}
            \label{subfig:noise_36}
        \end{subfigure}\hfill
        \begin{subfigure}{.19\linewidth}
            \includegraphics[width=\linewidth]{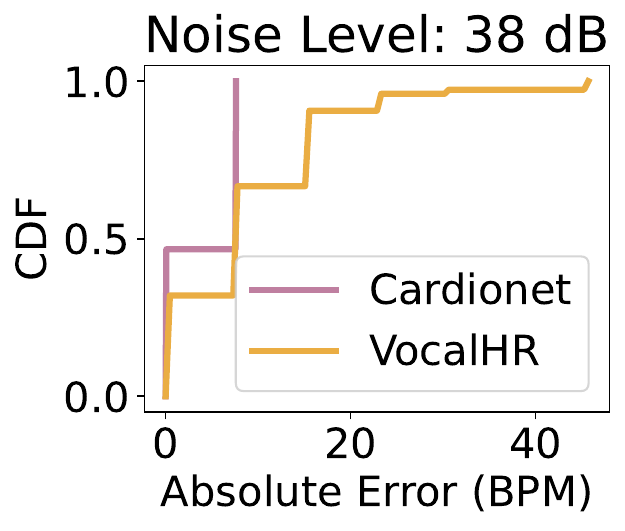}
            \label{subfig:noise_38}
        \end{subfigure}
    \end{minipage}
    \caption{CDF for different noise levels.}
    \label{fig: res_noise1}
\end{figure}

\begin{figure}[t]
    \centering
    \begin{minipage}{\linewidth}
        \begin{subfigure}{.325\linewidth}
            \includegraphics[width=\linewidth]{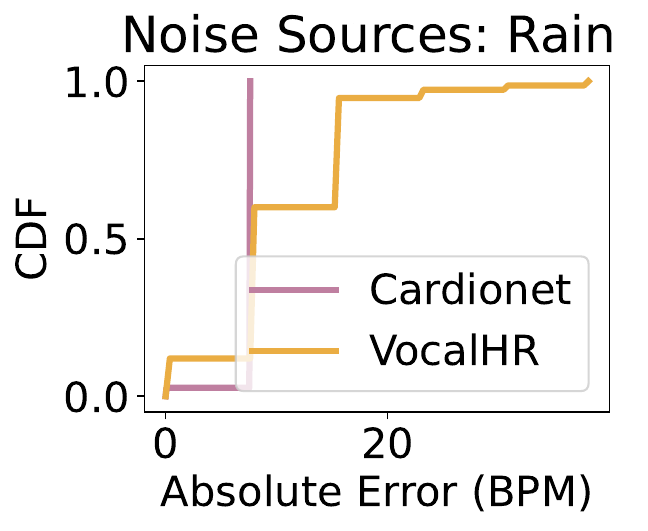}
            \label{subfig:noise_rain}
        \end{subfigure}\hfill
        \begin{subfigure}{.325\linewidth}
            \includegraphics[width=\linewidth]{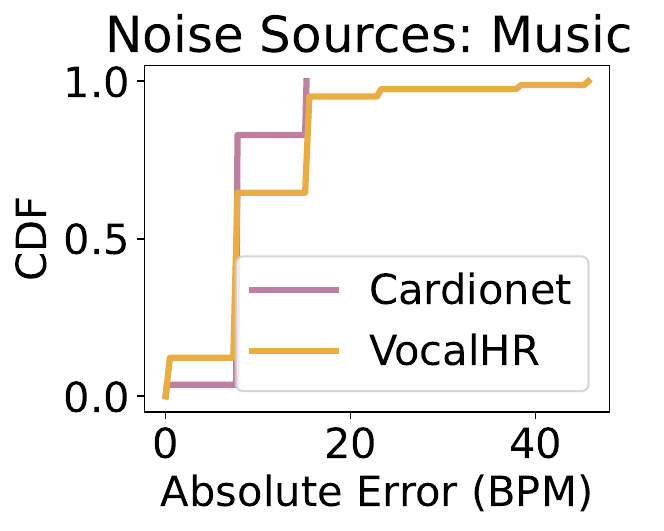}
            \label{subfig:noise_music}
        \end{subfigure}\hfill
        \begin{subfigure}{.34\linewidth}
            \includegraphics[width=\linewidth]{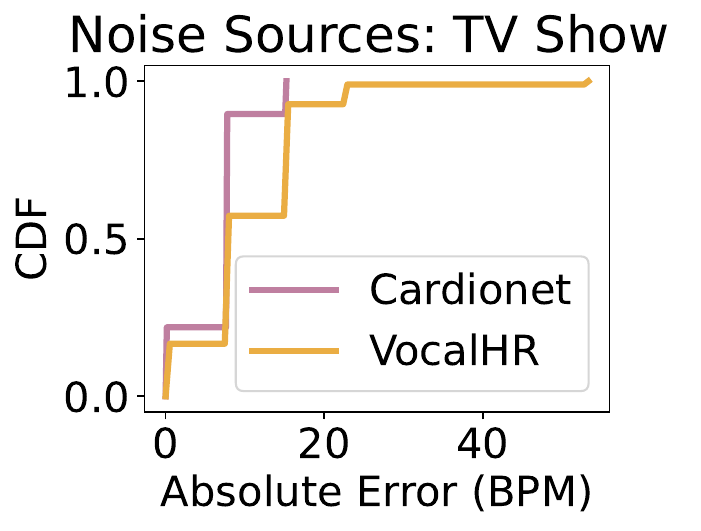}
            \label{subfig:noise_tv}
        \end{subfigure}
    \end{minipage}
    \caption{CDF for different noise sources.}
    \label{fig: res_noise2}
\end{figure}

\head{Noise Levels}
We test heart rate under noise levels from 30 dB to 38 dB. As in \fig\ref{fig: res_noise1}, increasing noise leads to higher error. Nonetheless, \texttt{CardioNet} consistently outperforms the SOTA audio-only model \texttt{VocalHR}. This can be attributed to our temporal frequency filter design and the video modality, which provides complementary information that remains stable under acoustic noise.  For instance, at 30 dB, \sysname achieves a MAE of 1.25 BPM, significantly lower than \texttt{VocalHR}'s 8.64 BPM, and maintains this advantage even at 38 dB. 
The fusion network learns to adaptively reduce reliance on noisy audio features while keeping stable visual cues. 
The CDF curves show that \texttt{CardioNet} achieves higher cumulative probabilities at lower error thresholds, indicating its resilience to noise.

\head{Noise Sources}
We analyze the impact of noise sources such as rain, music, and TV shows in \fig\ref{fig: res_noise2}. \texttt{CardioNet} demonstrates strong noise resilience, particularly with rain noise, where it significantly outperforms VocalHR, achieving a MAE of just 1.94 BPM compared to 12.93 BPM. 
Even with more complex noise like music and TV shows, our model maintains lower MAEs, showcasing its robustness in diverse acoustic environments. This highlights the effectiveness of video modalities when facing ambient noise.

\begin{figure}[t]
    \centering
    \begin{minipage}{\linewidth}
        \centering
        \includegraphics[width=\linewidth]{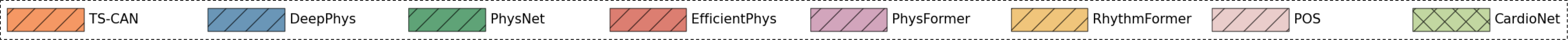}
    \end{minipage}
    
    \begin{minipage}{\linewidth}
        \begin{subfigure}{.32\linewidth}
            \includegraphics[width=\linewidth]{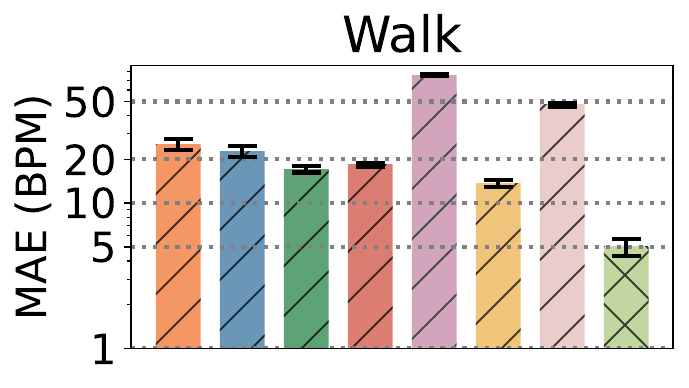}
            \caption{Walk}
            \label{subfig:move_walk}
        \end{subfigure}\hfill
        \begin{subfigure}{.32\linewidth}
            \includegraphics[width=\linewidth]{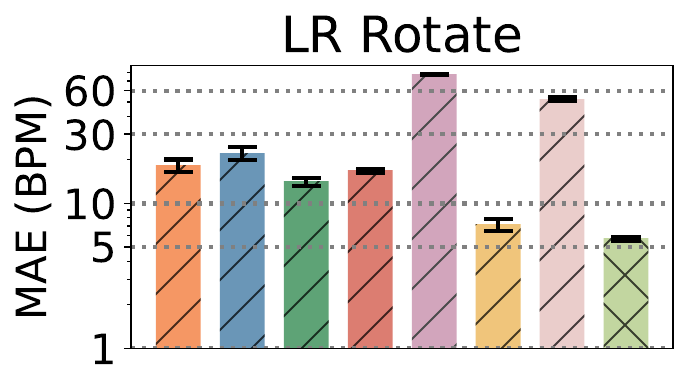}
            \caption{Left-Right}
            \label{subfig:move_lr}
        \end{subfigure}\hfill
        \begin{subfigure}{.32\linewidth}
            \includegraphics[width=\linewidth]{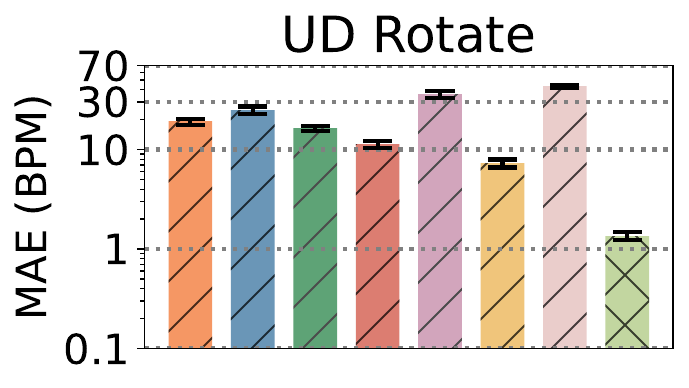}
            \caption{Up-Down}
            \label{subfig:move_ud}
        \end{subfigure}
    \end{minipage}
    \caption{The performances of different body motions.}
    \label{fig: res_move}
\end{figure}

\head{Body Motions}
Body motion can significantly impact the performance of heart rate detection models. To validate the robustness of our approach, we evaluate the model in three typical body movements: walking, left-right (LR) rotation, and up-down (UD) rotation, as in \fig\ref{fig: res_move}. 
Despite the motion artifacts, \texttt{CardioNet} maintains robust performance, achieving an MAE of 1.35 BPM in the UD scenario, and consistently outperforms baselines by significant margins in all motion types. Our model benefits from the unique design of the motion-aware aggregation and temporal differentiation block.
These prove the robustness of our model against body motions by effectively employing video plus audio modalities.

\begin{figure}[t]
    \centering

    \begin{minipage}{0.8\linewidth}
        \centering
        \includegraphics[width=\linewidth]{pdf/videoonly_legend.pdf}
    \end{minipage}

    \begin{minipage}{0.8\linewidth}
        \begin{subfigure}{.51\linewidth}
            \includegraphics[width=\linewidth]{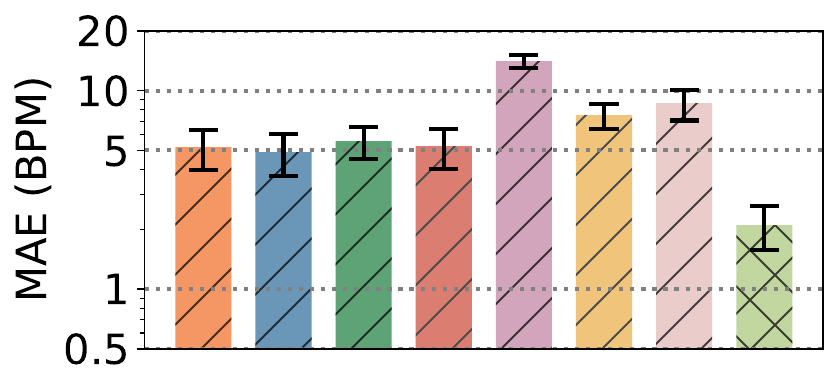}
            \caption{PURE}
            \label{subfig:res_pure}
        \end{subfigure}\hfill
        \begin{subfigure}{.49\linewidth}
            \includegraphics[width=\linewidth]{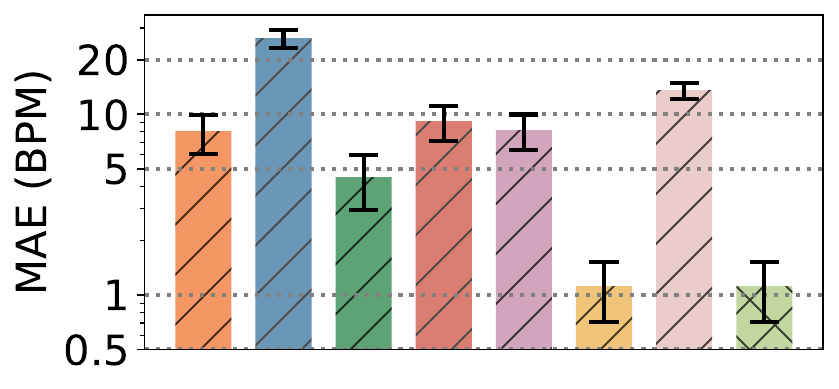}
            \caption{MMPD}
            \label{subfig:res_mmpd}
        \end{subfigure}\hfill
    \end{minipage}
    \caption{The performances on public datasets.}
    \label{fig: res_pure}
\end{figure}

\head{Video-only Solutions} We evaluate our approach on open datasets that contain only video data. As shown in \fig\ref{fig: res_pure}, our method consistently ranks among the top among rPPG-based solutions. We achieve MAE errors of 2.09 and 1.12 BPM on PURE and MMPD datasets, respectively.
It is important to note that during evaluation, we disable the audio branch of \texttt{CardioNet}. This ensures that our video encoder independently captures heart-related activities. 
In scenarios where no audio is available (e.g., during silent periods), our model effectively transitions into a video-only solution.

\tmc{

\head{Cross-Dataset Performance} To validate the generalizability of the model, we perform cross-dataset experiments on the current SOTA video-only solution and ours. As shown in \tab\ref{tab:model_comparison}, we find that our model is significantly better when training on PURE and testing on MMPD, with an MAE of 2.845 BPM. This is because MMPD is a complicated dataset, where the motion and the light varies a lot. This can also be seen from \fig\ref{fig: mmpd}, which shows that when confronted with different motions and lights, previous methods fail to provide a robust way to handle them. Conversely, our method incorporates motion-aware and frequency-aware modeling, which will enhance the performance.
Furthermore, to our knowledge, no previous work has reported results for training on MMPD and testing on PURE. Our method achieves an MAE of 3.675 BPM and an RMSE of 8.07 in this setting. These results demonstrate the generalizability of our approach across different scenarios.
\begin{table}[h]
    \centering
    \caption{\tmc{Cross-Dataset Performances on PURE and MMPD datasets.}}
    \label{tab:model_comparison}
    \colortable{
    \begin{tabular}{lccrr}
        \toprule
        \textbf{Model} & \textbf{Train-Set} & \textbf{Test-Set} & \textbf{MAE} & \textbf{RMSE} \\
        \midrule
        TS-CAN \cite{liu2020multi}       & PURE  & MMPD & 13.93   & 15.14   \\
        PhysNet \cite{yu2019remote}      & PURE  & MMPD & 13.93   & 15.61   \\
        PhysFormer \cite{yu2023physformer++}   & PURE  & MMPD & 14.57   & 16.73   \\
        DeepPhys \cite{chen2018deepphys}     & PURE  & MMPD & 16.92   & 18.54   \\
        EfficientPhys \cite{liu2023efficientphys} & PURE  & MMPD & 14.03   & 15.31   \\
        RhythmFormer \cite{zou2024rhythmformer}  & PURE  & MMPD & 8.98    & 14.85   \\
        \midrule
        \textbf{Ours} & PURE  & MMPD & \textbf{2.845}  & \textbf{6.688} \\
        \midrule
        \textbf{Ours} & MMPD  & PURE & \textbf{3.675}  & \textbf{8.07}  \\
        \bottomrule
    \end{tabular}
    }
\end{table}

}


\begin{table}[ht]
    \centering
    \caption{\tmc{Ablation study evaluating contributions of each component.
    ``w/o'' denotes ablative removal; ``w/o raw audio'' denotes replacing raw audio input with Mel-Spectrogram.}}
    \label{tab:ablation_study}
    \colortable{
    \begin{tabular}{lccc}
        \toprule
        & MAE & RMSE & MAPE \\
        \midrule
        w/o audio & 8.400 $\pm$ 0.284 & 8.898 $\pm$ 5.810 & 0.113 $\pm$ 0.004 \\
        w/o ITE & 3.168 $\pm$ 0.388 & 5.028 $\pm$ 3.486 & 0.045 $\pm$ 3.486 \\
        w/o FCB & 3.660 $\pm$ 0.456 & 5.421 $\pm$ 4.212 & 0.048 $\pm$ 0.006 \\
        w/o FCB + w/o ITE & 4.353 $\pm$ 0.473 & 6.014 $\pm$ 4.888 & 0.063 $\pm$ 0.007 \\
        w/o Raw Audio & 4.168 $\pm$ 0.221 & 5.914 $\pm$ 2.314 & 0.061 $\pm$ 0.003 \\
        \midrule
        \textbf{Ours (w/ all)} & \textbf{1.746 $\pm$ 0.380} & \textbf{4.114 $\pm$ 4.516} & \textbf{0.024 $\pm$ 0.005} \\
        \bottomrule
    \end{tabular}
    }
\end{table}
\tmc{

\subsection{Ablation Study}

To validate the contributions of each component in the model, we perform a comprehensive ablation study as shown in \tab\ref{tab:ablation_study}.

\head{w/o Audio} We first evaluate the model’s performance without the audio modality. As shown in the results, the MAE increases sharply from 1.746 to 8.400, indicating a significant performance drop. This demonstrates that audio–video fusion is highly effective, and that our temporal attention fusion mechanism successfully leverages complementary cues from both modalities.

\head{w/o Irregular Time Embedding (ITE)} The Irregular Time Embedding (ITE) component is designed to address the irregular sampling inherent in real-world video data. Removing this component results in a performance deterioration of 81.4\% in MAE, indicating that temporal embedding is essential for handling unconstrained video streams.

\head{w/o Frequency-Aware Conv Block (FCB)} The Frequency-Aware Conv Block (FCB) is specifically designed to enhance the model’s ability to capture subtle frequency information and to aggregate it with spatial representations. Excluding FCB leads to a 52.3\% increase in MAE, confirming its positive impact on model accuracy.

\head{w/o ITE + w/o FCB} When both ITE and FCB are simultaneously removed, model performance degrades further compared to removing either module alone. This suggests that ITE and FCB independently and jointly enhance the model’s ability to process complex signals.

\head{w/o Raw Audio} As discussed in \S\ref{sec: cardionet}, we propose using raw audio instead of the Mel-Spectrogram based on empirical observations. In this experiment, we replace the raw audio input with features extracted from Mel-Spectrograms using a ResNet encoder. The MAE increases from 1.746 BPM (raw audio) to 4.168~BPM (Mel-Spectrogram), confirming that raw audio is a more effective input representation for our task. Nonetheless, the inclusion of Mel-Spectrogram features still yields better results than removing the audio modality entirely, further highlighting the importance of audio in video-based cardiac monitoring systems compared to video-only solutions.

}

\subsection{Micro Benchmarks}

\begin{figure}[t]
    \centering
    \begin{minipage}{\linewidth}
        \begin{subfigure}{0.32\linewidth}
            \includegraphics[width=\linewidth]{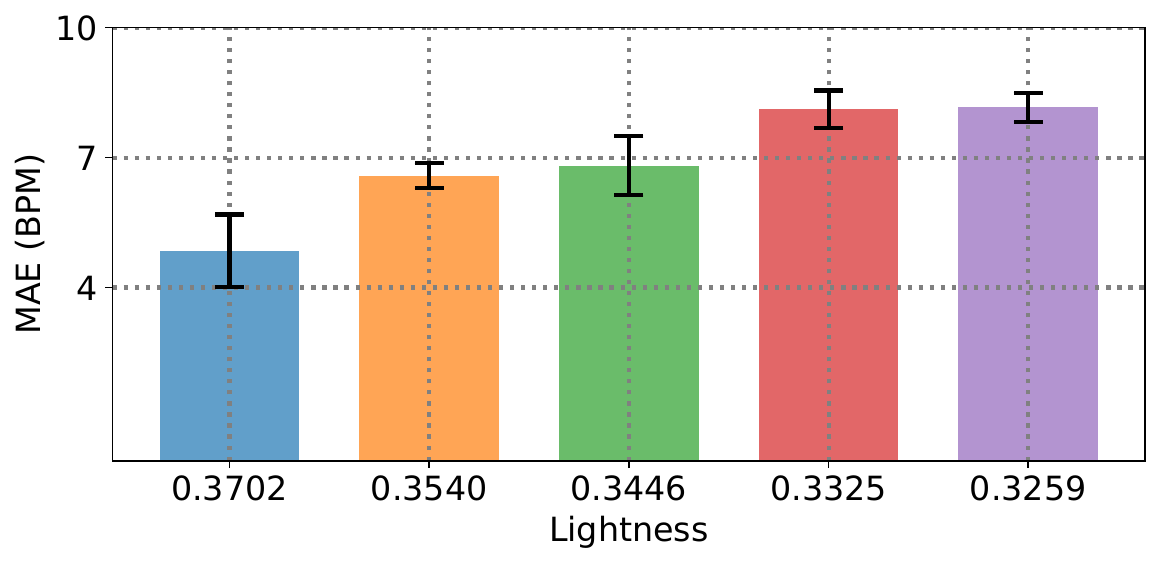}
            \caption{Lightness}
            \label{subfig:light}
        \end{subfigure}\hfill
        \begin{subfigure}{0.32\linewidth}
            \includegraphics[width=\linewidth]{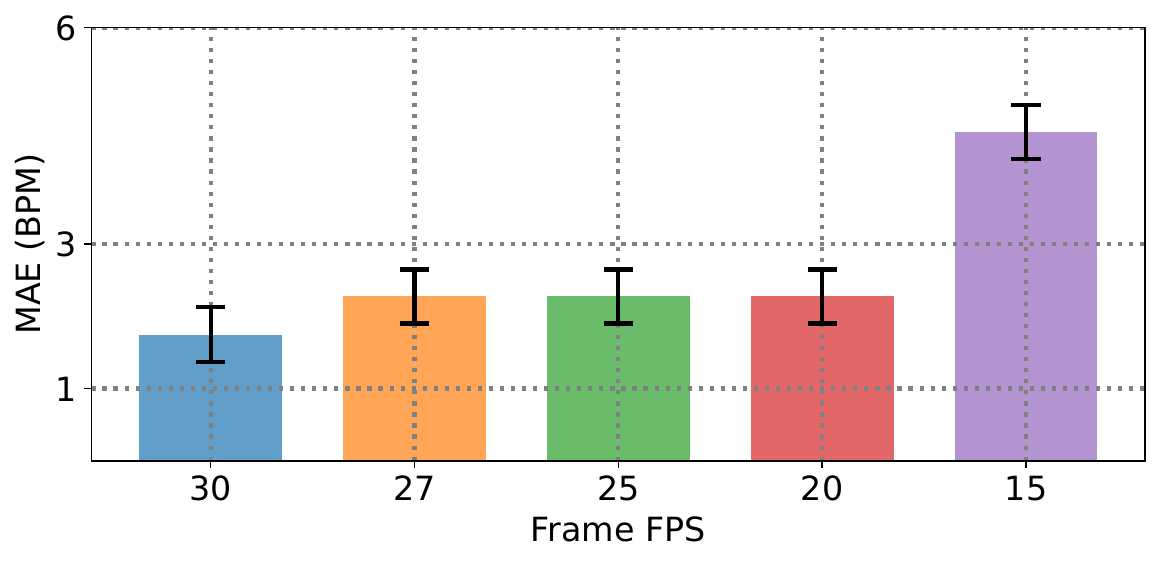}
            \caption{FPS}
            \label{subfig:FPS}
        \end{subfigure}\hfill
        \begin{subfigure}{0.32\linewidth}
            \includegraphics[width=\linewidth]{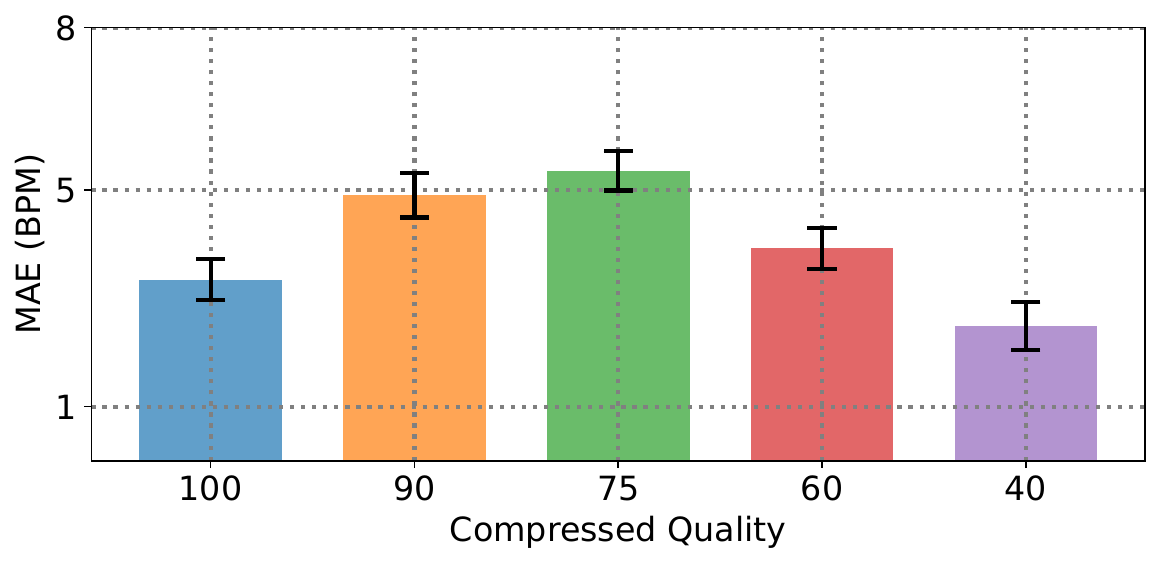}
            \caption{QoI}
            \label{subfig:qoi}
        \end{subfigure}
        \caption{The performances under different light, FPS and quality of image.}
        \label{fig: res_light-FPS-qoi}
    \end{minipage}\hfill
\end{figure}

\head{Different Light Conditions}
We assess our model under varying lightness levels from 0.3702 to 0.3259 in Fig.\ref{subfig:light}, by adjusting the ring light. As it decreases, the MAE increases from 4.85 BPM to 8.16 BPM. This trend suggests that poorer conditions impact accuracy due to the reduced visibility of facial features. However, the model remains sufficiently robust, indicating that while lighting plays a role, the audio-visual fusion helps mitigate the negative effects.

\head{Different FPS}
We examine the model across various video frame rates, ranging from 30 to 15 FPS, as shown in Fig.\ref{subfig:FPS}.  We interpolate the frame rate by adopting the principles discussed in \S\ref{subsec: service}. The model performs best at 30 FPS with an MAE of 1.75 BPM. Even at lower frame rates, particularly 15 FPS, the MAE increases to 4.56 BPM, while still remaining in a low level.  This performance is achieved through our frame interpolation scheme and the audio branch's ability to provide continuous cardiac information. Also, our temporal differential block and irregularly sampled time embedding block are equally vital to handle varying frame rates.

\head{Different Quality of Image}
We analyze the performance under various video compression qualities, from 100 to 40 (lowest quality), as shown in Fig.\ref{subfig:qoi}. 
The MAE does not consistently worsen with lower quality. At extreme compression levels, the model achieves the lowest MAE of 2.49 BPM, potentially due to smoothing effects that enhance key facial features. 
This suggests that while high compression degrades visual information, moderate to high levels of compression might benefit the model by reducing noise.

\tmc{
\begin{figure}[t]
\centering
    \begin{minipage}{0.8\linewidth}
        \begin{subfigure}{0.42\linewidth}
            \includegraphics[width=\linewidth]{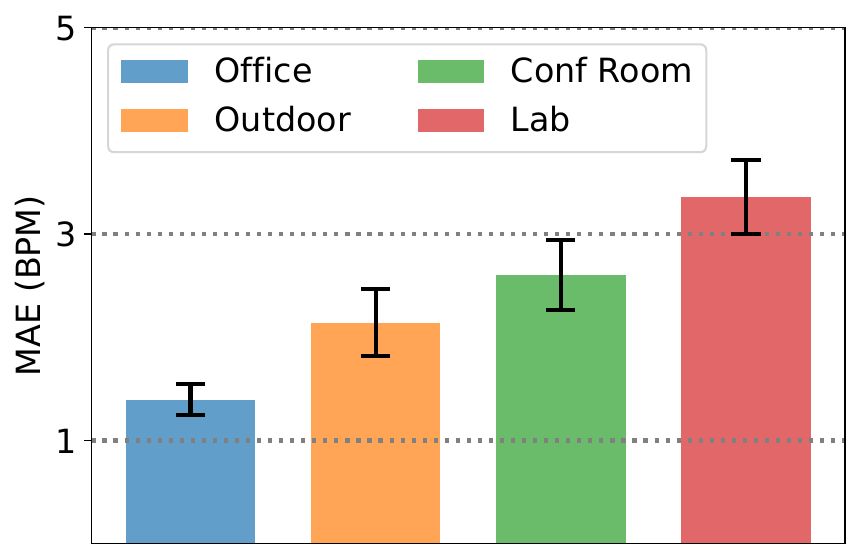}
            \caption{\tmc{Environment}}
            \label{subfig:env}
        \end{subfigure}\hfill
        \begin{subfigure}{.57\linewidth}
            \includegraphics[width=\linewidth]{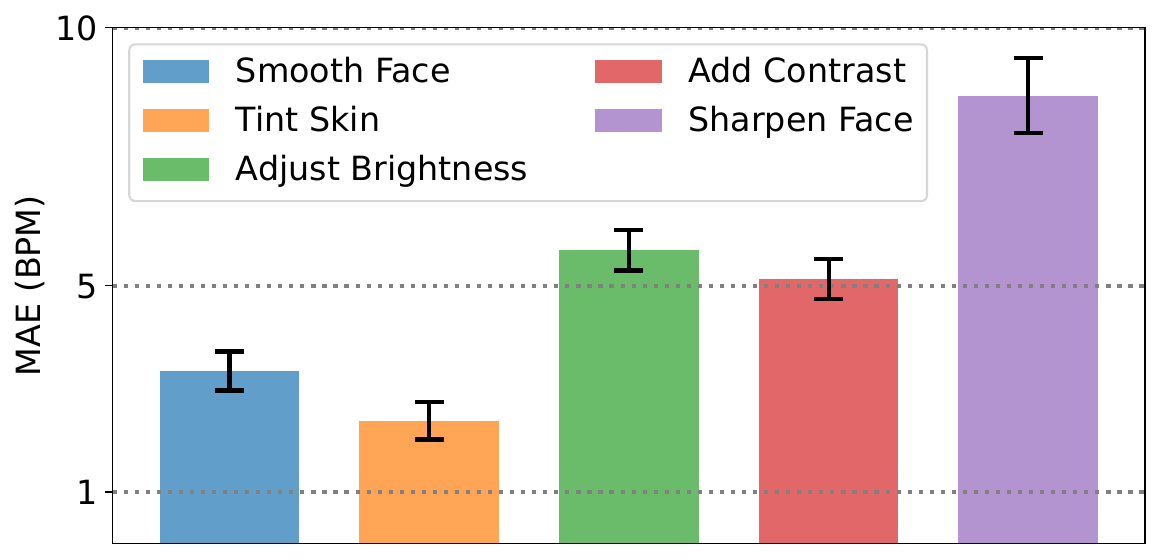}
            \caption{\tmc{Face Beauty}}
            \label{subfig:fb}
        \end{subfigure}\hfill
    \end{minipage}
    \caption{\tmc{The performances under different environments and face beauty filters.}}
    \label{fig: res_fb-env}
\end{figure}
}
\tmc{
\head{Different Environments}
Our model's performance is evaluated across various environmental settings, including Office, Outdoor, Conference Room, and Laboratory, as shown in Fig.\ref{subfig:env}. 
The model performs best in the Office environment with an MAE of 1.40 BPM. Notably, the latter three environments are not in the training set, yet the model maintains strong performance, demonstrating that our feature extraction generalizes well to unseen conditions.

\head{Different Face Filters}
We test various facial filters, including Smooth Face, Tint Skin, Adjust Brightness, Add Contrast, and Sharpen Face, as shown in Fig.\ref{subfig:fb}. 
The Tint Skin filter yields the best performance with an MAE of 2.38 BPM, while a more aggressive filter like Sharpen Face achieves an MAE of 8.69 BPM. It shows our model effectively handles appearance changes.
}

\tmc{
\head{Different Devices}
We evaluate our model on various devices under inter-device and cross-device conditions, as shown in Fig.\ref{subfig:device}.
For inter-device testing, the average MAE is approximately 2.95 BPM. In cross-device scenarios, the average MAE is around 8.07 BPM. While there is a drop in accuracy, the model still delivers acceptable performance across different hardware platforms. 
This suggests that despite some variability, the model remains robust and capable of providing reliable heart rate estimates on a wide range of devices. 

\head{Different Users}
We evaluate our model's performance across a diverse set of users in Fig.\ref{subfig:people}. Our model's user generalization capability stems from learning universal cardiac patterns rather than user-specific features. The temporal-spectral modeling captures fundamental physiological characteristics that are consistent across individuals. Under inter-user conditions, the average MAE is about 1.93 BPM. In cross-user scenarios, the model still performs reasonably well, with an average MAE of 7.53 BPM. Despite the diversity, the model maintains a usable level of accuracy, underscoring its generalizability across different user groups. This demonstrates that our feature extraction pipeline effectively captures device-independent cardiac patterns.

\head{Multi-person Scenarios} We evaluate the multi-person scenarios to justify the effectiveness of our preprocessing. 
We set the maximum number of people to be separated as two and crop the face region to a size of 72$\times$72 pixels. 
In our test, two users read materials simultaneously while sitting next to each other. We apply the facial and sound separation and match their audio and face regions. The test results show an MAE of 7.83 BPM and 8.13 BPM for each person, respectively. 
Although we observe some performance drops, our method still effectively distinguishes between the two individuals. Notably, the heart rates of the two people vary over time, with average heart rates of 76.17 BPM and 68.55 BPM, respectively, showing our system can track distinct physiological states simultaneously.

\begin{figure}[t]
    \centering
    \begin{minipage}{0.5\linewidth}
         \includegraphics[width=\linewidth]{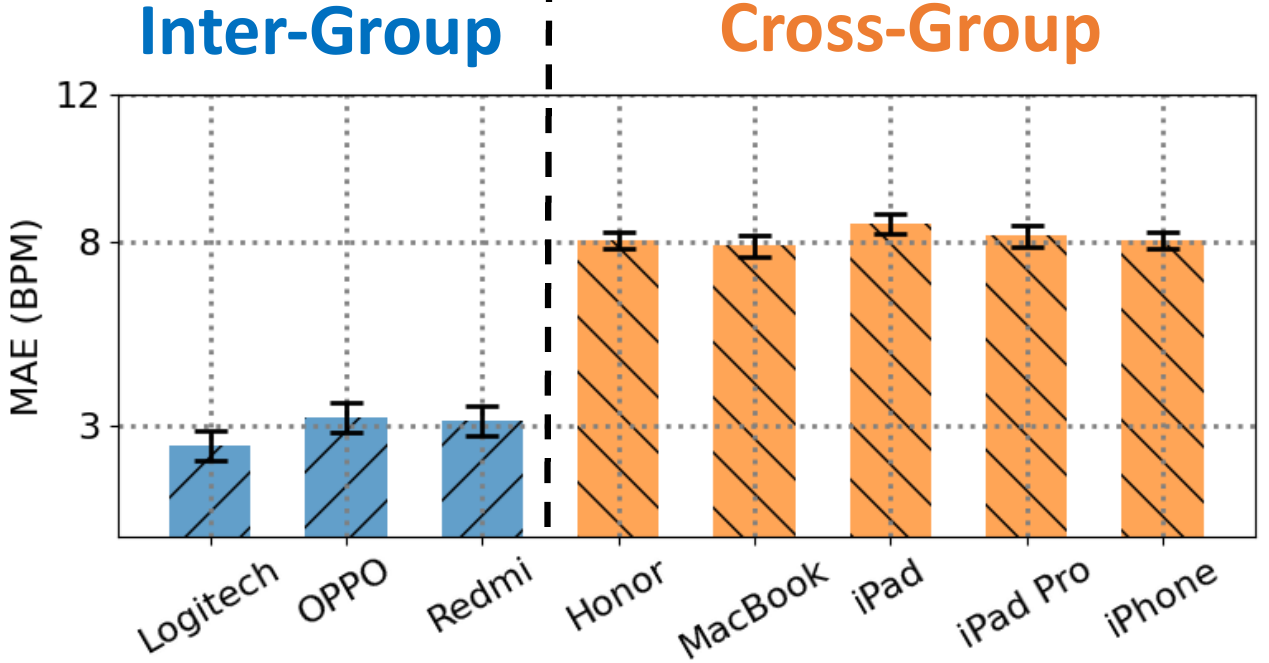}
        \caption{\tmc{Different Devices}}
        \label{subfig:device}
    \end{minipage}\hfill
    \begin{minipage}{0.49\linewidth}
        \includegraphics[width=\linewidth]{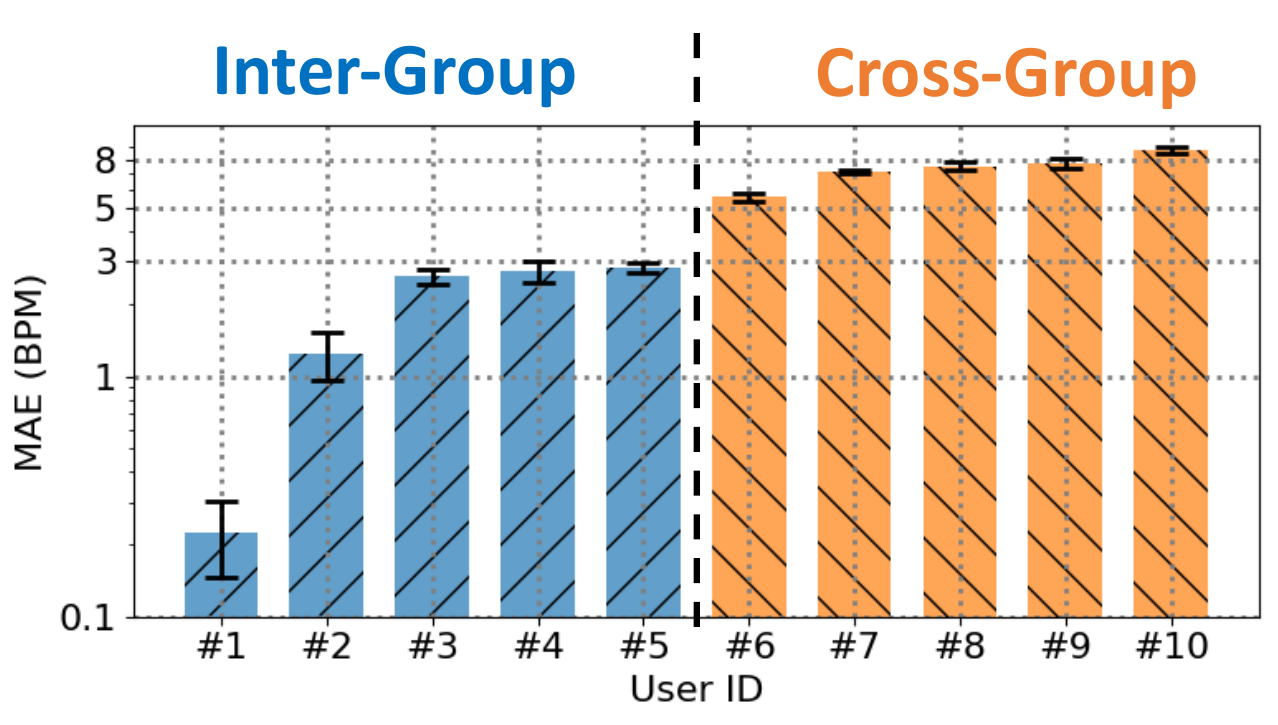}
        \caption{\tmc{Different Users}}
        \label{subfig:people}
    \end{minipage}\hfill
\end{figure}

}

\subsection{\sysname in the wild}
In this section, we will evaluate how \sysname works as a service.

\begin{figure}[t]
    \centering

    \begin{minipage}{\linewidth}
        \begin{subfigure}{0.49\linewidth}
            \includegraphics[width=\linewidth]{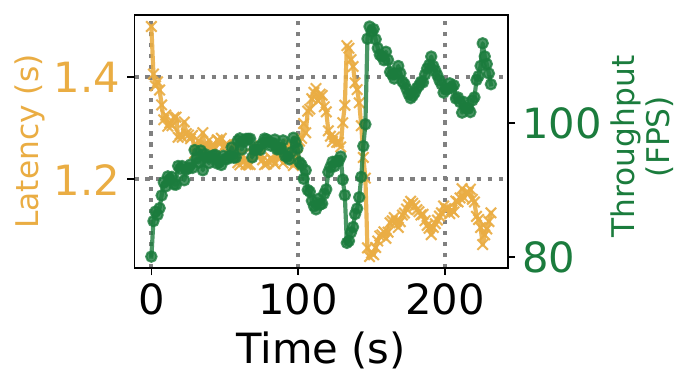}
            \caption{Chrome}
            \label{subfig:inference_chrome}
        \end{subfigure}\hfill
        \begin{subfigure}{0.49\linewidth}
            \includegraphics[width=\linewidth]{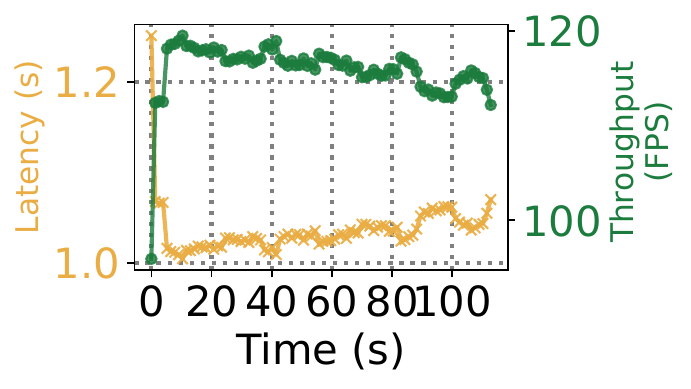} %
            \caption{Zoom}
            \label{subfig: inference_zoom}
        \end{subfigure}
    \end{minipage}
    \caption{Latency \& throughput for Zoom and Chrome}
    \label{fig: res_througput}
\end{figure}

\head{Meeting Platforms}
We choose Zoom as one of the online meeting platforms, which provides the external developers with the SDK to acquire access to the raw data. The average FPS is 28.4. We exploit the data hooks to acquire the streams and leverage buffer queues to hold the packets, as described in \S\ref{subsec: buffer}. The model consumes on average in 850ms on CPU with a step size of 1s and an inference window size of 4s. The overall system latency averages 1.03 seconds, as depicted in \fig\ref{subfig: inference_zoom}. Notably, latency was primarily elevated at the start due to the initial model warm-up period \cite{lion2016don}. This means our systems can run inference in real-time.
Furthermore, we calculate the throughput of the whole system. We measure the time since the last update of the heart rate. As we are feeding a 4-s window of video and audio frames, the throughput is calculated as the volume of video and audio data processed per update period. As in \fig\ref{subfig: inference_zoom}, the average throughput of the system is 115.97 FPS, which is prominently larger than the common video FPS. It means that our systems can hold the service robustly without any freezes.

\head{Online Content Providers}
Online content providers such as YouTube often host their services in the web browser. We implement such a service in a Chrome extension. We employ the data hook to acquire the streams. The average FPS is 26.97. The overall latency of our service is 1.23s, comparable to our step size 1s, as can be observed from \fig\ref{subfig:inference_chrome}. Meanwhile, the average throughput is 98.16 FPS, with a maximum throughput of 114.41 FPS. These results also justify our service will run smoothly in the extensions.

\tmc{
\head{Model Size} Our model contains 81.58M parameters and requires 7.398G FLOPs, which are comparable to other audio-video learning frameworks \cite{gong2022contrastive}. We have further pruned and quantized the model for faster inference. Note that we must wait for the first window before processing; however, this initial latency is standard and acceptable for this type of task in the literature.
}

\section{Related Work}
In this section, we will summarize the existing works.

\head{Cardiac Monitoring}
Cardiac information is crucial for health monitoring, affective computing \cite{yang2022survey, fairclough2020personal} and deception analysis \cite{bian2024ubihr}. 
Compared with hospital solutions \cite{HeartDiseaseSymptoms}, recent advancements have focused on more portable solutions \cite{xue2025ecg, alimbayeva2022portable, chan2025mobile, chan2019contactless}.
Earable systems \cite{cao2023heartprint, chen2024exploring, fan2023apg, cao2024earsteth, bui2019ebp, park2020heartquake} allow earpieces to detect cardiac information, but they either need specific probing signals or custom hardware, limiting their widespread adoption. Similarly, wearable solutions necessitate constant wear, which is not practical for all users. Wireless technologies, including Wi-Fi \cite{liu2015tracking}, mmWave \cite{yang2016monitoring}, and UWB \cite{chen2021movi}, \etc, are constrained by specific hardware that is not commonly available in video systems. 
Solutions using active acoustic sensing \cite{wang2023df, wang2022loear, qian2018acousticcardiogram, zhang2020your} with smart speakers rely on pseudo-inaudible signals, which can be intrusive to human hearing and increase hardware burden.
Video-based solutions use optical means to measure blood volume changes in tissues. Signal processing \cite{de2013robust, li2014remote, wang2016algorithmic, wang2015novel} and deep learning \cite{chen2018deepphys, liu2020multi, niu2020video, li2023learning, yu2019remote,yu2019remoteCompress, yu2023physformer++, liu2023efficientphys, zou2024rhythmformer, qian2024cluster, wang2024rppg, zou2025rhythmmamba, luo2024physmamba, wu2025cardiacmamba} techniques have been developed to enhance these methods. Yet these solutions are sensitive to low light conditions, head/body movements, and typically perform poorly outside controlled environments. VocalHR \cite{xu2022hearing} proves the potential of extracting heart rate from human speech. However, it is limited by range and requires pre-calibration. 
Differently, \sysname is the first to combine the complementary and naturally co-existing audio and video modalities in online video streaming systems. Our video design incorporates temporal-frequency co-design and motion-aware aggregations for the first time in OCM to mitigate the light and body movement influence. The audio module employs the temporal acoustic filter for OCM. These designs are innovative and contribute to our performances.

\head{Video Streaming System} Video streaming systems have gained immense popularity due to their vast libraries of on-demand content, user-generated videos, and live streaming capabilities, catering to diverse viewer preferences, including YouTube, TikTok, Zoom, \etc. They can be further categorized into VoD systems, live streaming systems and video conferencing systems. Research efforts have been devoted to communication protocols \cite{hamadanian2023ekho, dhawaskar2023converge}, adaptive rate streaming algorithms  \cite{li2023dashlet, wen2023adaptivenet, zhou2019learning, li2022livenet, tashtarian2024artemis}, online learning \cite{tang2023lut, guan2023metastream, khani2023recl, yi2023boosting, zhang2024aralive, wu2024toward, chen2024videollm, qian2024streaming}, and video understanding and serving \cite{li2025videoscan, tong2022videomae, naiman2025lv, du2020server}, \etc. None of these works explores adding cardiac monitoring to modern video streaming systems. 
In contrast, \sysname stands out as the first work that creates a middleware service of OCM that can be seamlessly integrated into mainstream video streaming systems.

\section{Discussion and Future Work}

\tmc{
\head{Audio-Video Pair} In our primary application scenarios (\eg, live streaming, online meetings, \etc), audio and video naturally coexist. 
In practice, only video data is available in some situations, where \sysname can be easily adapted to a video-only solution. Such periods can be detected through mature voice activity detection techniques \cite{wiseman2016python}. 
Our results shown in \fig\ref{fig: res_pure} have demonstrated that \sysname also performs well in video-only scenarios. 
\sysname not only introduces a novel approach to OCM by utilizing audio-visual pairs for the first time, but also integrates these capabilities into a practical system with flexibility and robustness.

}

\head{Impacts on Original Streams} Integrating additional services into streaming platforms can be a bottleneck for many previous solutions \cite{liu2020grad, mmdetection, du2020server}. In \sysname, we address it with a dedicated design of data hook and middleware service. 
Our approach ensures that these additional services are isolated from the original streams.
With an offscreen canvas, we avoid disrupting the original content. In meetings, our data hook duplicates data to the inference engine instantly without affecting the main video and audio streams. Our evaluations demonstrate that \sysname operates without causing any disruptions or interference to ongoing streams.

\head{Equality and Accessibility} \sysname is designed for equality and is devised to be flexible and adaptable, allowing it to be integrated into any platform without the need for specialized hardware. This significantly increases accessibility, making the technology available to a wider audience. Moreover, while companies can promote this service on cloud platforms, \sysname is crafted to ensure democratized access, preventing any hidden biases or preferential treatment. By enabling audiences to independently initiate the service, \sysname reduces the likelihood of companies manipulating the system for economic gains by altering the model.

\tmc{
\head{Use of Deep Learning} The relationship between video-audio information and cardiac activity is inherently implicit and complex. We evaluate our results against signal processing approaches in \fig\ref{fig: res_dis} and \fig\ref{fig: res_angle}, where our performances are significantly better. And our system evaluation validates real-time monitoring without introducing large latency. We identify the exploration of combining signal processing with increased explainability as a direction for future work.
}

\section{Conclusion}
In this paper, we envision the attractiveness of Online Cardiac Monitoring (OCM) in video streaming 
and present \sysname, the first system to fuse both audio and video streams for OCM. We devise an effective audio-visual network that can robustly and accurately unveil the nuanced cardiac activities, achieving an average MAE of 1.79 BPM and outperforming the video-only and audio-only solutions by 69.2\% and 81.2\%, respectively. Furthermore, we design and implement \sysname as a plug-and-play middleware that can seamlessly be integrated into mainstream streaming systems. We believe our work will significantly enhance the entertainment and healthcare value of video streaming and inspire new directions.

\section*{Acknowledgments}
This work is supported by the NSFC under Grant No. 62222216, the Hong Kong RGC ECS under Grant No. 27204522, and GRF under Grant No. 17212224.

\bibliographystyle{IEEEtran}
\bibliography{refs}

@misc{HeartDiseaseSymptoms,
	title = {Heart disease - {Symptoms} and causes - {Mayo} {Clinic}},
	url = {https://www.mayoclinic.org/diseases-conditions/heart-disease/symptoms-causes/syc-20353118},
	urldate = {2024-07-10},
}

@misc{PulsoidRealtimeHearta,
	title = {Pulsoid - a real-time heart rate widget for streaming},
	url = {https://pulsoid.net/},
	urldate = {2024-08-26},
}

@inproceedings{xu2022hearing,
  title={Hearing heartbeat from voice: Towards next generation voice-user interfaces with cardiac sensing functions},
  author={Xu, Chenhan and Chen, Tianyu and Li, Huining and Gherardi, Alexander and Weng, Michelle and Li, Zhengxiong and Xu, Wenyao},
  booktitle={Proceedings of the 20th ACM Conference on Embedded Networked Sensor Systems},
  pages={149--163},
  year={2022}
}

@inproceedings{liu2015tracking,
  title={Tracking vital signs during sleep leveraging off-the-shelf wifi},
  author={Liu, Jian and Wang, Yan and Chen, Yingying and Yang, Jie and Chen, Xu and Cheng, Jerry},
  booktitle={Proceedings of the 16th ACM international symposium on mobile ad hoc networking and computing},
  pages={267--276},
  year={2015}
}

@inproceedings{chen2021movi,
  title={MoVi-Fi: Motion-robust vital signs waveform recovery via deep interpreted RF sensing},
  author={Chen, Zhe and Zheng, Tianyue and Cai, Chao and Luo, Jun},
  booktitle={Proceedings of the 27th annual international conference on mobile computing and networking},
  pages={392--405},
  year={2021}
}

@inproceedings{yang2016monitoring,
  title={Monitoring vital signs using millimeter wave},
  author={Yang, Zhicheng and Pathak, Parth H and Zeng, Yunze and Liran, Xixi and Mohapatra, Prasant},
  booktitle={Proceedings of the 17th ACM international symposium on mobile ad hoc networking and computing},
  pages={211--220},
  year={2016}
}

@inproceedings{cao2023heartprint,
  title={HeartPrint: Passive heart sounds authentication exploiting in-ear microphones},
  author={Cao, Yetong and Cai, Chao and Li, Fan and Chen, Zhe and Luo, Jun},
  booktitle={IEEE INFOCOM 2023-IEEE Conference on Computer Communications},
  pages={1--10},
  year={2023},
  organization={IEEE}
}

@inproceedings{wang2023df,
  title={Df-sense: Multi-user acoustic sensing for heartbeat monitoring with dualforming},
  author={Wang, Lei and Gu, Tao and Li, Wei and Dai, Haipeng and Zhang, Yong and Yu, Dongxiao and Xu, Chenren and Zhang, Daqing},
  booktitle={Proceedings of the 21st Annual International Conference on Mobile Systems, Applications and Services},
  pages={1--13},
  year={2023}
}

@article{wang2022loear,
  title={LoEar: Push the range limit of acoustic sensing for vital sign monitoring},
  author={Wang, Lei and Li, Wei and Sun, Ke and Zhang, Fusang and Gu, Tao and Xu, Chenren and Zhang, Daqing},
  journal={Proceedings of the ACM on Interactive, Mobile, Wearable and Ubiquitous Technologies},
  volume={6},
  number={3},
  pages={1--24},
  year={2022},
  publisher={ACM New York, NY, USA}
}

@inproceedings{qian2018acousticcardiogram,
  title={Acousticcardiogram: Monitoring heartbeats using acoustic signals on smart devices},
  author={Qian, Kun and Wu, Chenshu and Xiao, Fu and Zheng, Yue and Zhang, Yi and Yang, Zheng and Liu, Yunhao},
  booktitle={IEEE INFOCOM 2018-IEEE conference on computer communications},
  pages={1574--1582},
  year={2018},
  organization={IEEE}
}

@article{zhang2020your,
  title={Your Smart Speaker Can" Hear" Your Heartbeat!},
  author={Zhang, Fusang and Wang, Zhi and Jin, Beihong and Xiong, Jie and Zhang, Daqing},
  journal={Proceedings of the ACM on Interactive, Mobile, Wearable and Ubiquitous Technologies},
  volume={4},
  number={4},
  pages={1--24},
  year={2020},
  publisher={ACM New York, NY, USA}
}

@inproceedings{chen2024exploring,
  title={Exploring the Feasibility of Remote Cardiac Auscultation Using Earphones},
  author={Chen, Tao and Yang, Yongjie and Fan, Xiaoran and Guo, Xiuzhen and Xiong, Jie and Shangguan, Longfei},
  booktitle={Proceedings of the 30th Annual International Conference on Mobile Computing and Networking},
  pages={357--372},
  year={2024}
}

@inproceedings{fan2023apg,
  title={Apg: Audioplethysmography for cardiac monitoring in hearables},
  author={Fan, Xiaoran and Pearl, David and Howard, Richard and Shangguan, Longfei and Thormundsson, Trausti},
  booktitle={Proceedings of the 29th Annual International Conference on Mobile Computing and Networking},
  pages={1--15},
  year={2023}
}

@article{de2013robust,
  title={Robust pulse rate from chrominance-based rPPG},
  author={De Haan, Gerard and Jeanne, Vincent},
  journal={IEEE transactions on biomedical engineering},
  volume={60},
  number={10},
  pages={2878--2886},
  year={2013},
  publisher={IEEE}
}

@inproceedings{li2014remote,
  title={Remote heart rate measurement from face videos under realistic situations},
  author={Li, Xiaobai and Chen, Jie and Zhao, Guoying and Pietikainen, Matti},
  booktitle={Proceedings of the IEEE conference on computer vision and pattern recognition},
  pages={4264--4271},
  year={2014}
}

@article{wang2015novel,
  title={A novel algorithm for remote photoplethysmography: Spatial subspace rotation},
  author={Wang, Wenjin and Stuijk, Sander and De Haan, Gerard},
  journal={IEEE transactions on biomedical engineering},
  volume={63},
  number={9},
  pages={1974--1984},
  year={2015},
  publisher={IEEE}
}

@inproceedings{chen2018deepphys,
  title={Deepphys: Video-based physiological measurement using convolutional attention networks},
  author={Chen, Weixuan and McDuff, Daniel},
  booktitle={Proceedings of the european conference on computer vision (ECCV)},
  pages={349--365},
  year={2018}
}

@article{liu2020multi,
  title={Multi-task temporal shift attention networks for on-device contactless vitals measurement},
  author={Liu, Xin and Fromm, Josh and Patel, Shwetak and McDuff, Daniel},
  journal={Advances in Neural Information Processing Systems},
  volume={33},
  pages={19400--19411},
  year={2020}
}

@inproceedings{niu2020video,
  title={Video-based remote physiological measurement via cross-verified feature disentangling},
  author={Niu, Xuesong and Yu, Zitong and Han, Hu and Li, Xiaobai and Shan, Shiguang and Zhao, Guoying},
  booktitle={Computer Vision--ECCV 2020: 16th European Conference, Glasgow, UK, August 23--28, 2020, Proceedings, Part II 16},
  pages={295--310},
  year={2020},
  organization={Springer}
}

@inproceedings{li2023learning,
  title={Learning motion-robust remote photoplethysmography through arbitrary resolution videos},
  author={Li, Jianwei and Yu, Zitong and Shi, Jingang},
  booktitle={Proceedings of the AAAI Conference on Artificial Intelligence},
  volume={37},
  number={1},
  pages={1334--1342},
  year={2023}
}

@article{yu2019remote,
  title={Remote photoplethysmograph signal measurement from facial videos using spatio-temporal networks},
  author={Yu, Zitong and Li, Xiaobai and Zhao, Guoying},
  journal={arXiv preprint arXiv:1905.02419},
  year={2019}
}

@inproceedings{yu2019remoteCompress,
  title={Remote heart rate measurement from highly compressed facial videos: an end-to-end deep learning solution with video enhancement},
  author={Yu, Zitong and Peng, Wei and Li, Xiaobai and Hong, Xiaopeng and Zhao, Guoying},
  booktitle={Proceedings of the IEEE/CVF international conference on computer vision},
  pages={151--160},
  year={2019}
}

@article{yu2023physformer++,
  title={Physformer++: Facial video-based physiological measurement with slowfast temporal difference transformer},
  author={Yu, Zitong and Shen, Yuming and Shi, Jingang and Zhao, Hengshuang and Cui, Yawen and Zhang, Jiehua and Torr, Philip and Zhao, Guoying},
  journal={International Journal of Computer Vision},
  volume={131},
  number={6},
  pages={1307--1330},
  year={2023},
  publisher={Springer}
}

@inproceedings{liu2023efficientphys,
  title={Efficientphys: Enabling simple, fast and accurate camera-based cardiac measurement},
  author={Liu, Xin and Hill, Brian and Jiang, Ziheng and Patel, Shwetak and McDuff, Daniel},
  booktitle={Proceedings of the IEEE/CVF winter conference on applications of computer vision},
  pages={5008--5017},
  year={2023}
}

@article{zou2024rhythmformer,
  title={Rhythmformer: Extracting rppg signals based on hierarchical temporal periodic transformer},
  author={Zou, Bochao and Guo, Zizheng and Chen, Jiansheng and Ma, Huimin},
  journal={arXiv preprint arXiv:2402.12788},
  year={2024}
}

@inproceedings{li2023dashlet,
  title={Dashlet: Taming swipe uncertainty for robust short video streaming},
  author={Li, Zhuqi and Xie, Yaxiong and Netravali, Ravi and Jamieson, Kyle},
  booktitle={20th USENIX Symposium on Networked Systems Design and Implementation (NSDI 23)},
  pages={1583--1599},
  year={2023}
}

@inproceedings{tang2023lut,
  title={Lut-nn: Empower efficient neural network inference with centroid learning and table lookup},
  author={Tang, Xiaohu and Wang, Yang and Cao, Ting and Zhang, Li Lyna and Chen, Qi and Cai, Deng and Liu, Yunxin and Yang, Mao},
  booktitle={Proceedings of the 29th Annual International Conference on Mobile Computing and Networking},
  pages={1--15},
  year={2023}
}

@inproceedings{hamadanian2023ekho,
  title={Ekho: Synchronizing cloud gaming media across multiple endpoints},
  author={Hamadanian, Pouya and Gallatin, Doug and Alizadeh, Mohammad and Chintalapudi, Krishna},
  booktitle={Proceedings of the ACM SIGCOMM 2023 Conference},
  pages={533--549},
  year={2023}
}

@inproceedings{guan2023metastream,
  title={Metastream: Live volumetric content capture, creation, delivery, and rendering in real time},
  author={Guan, Yongjie and Hou, Xueyu and Wu, Nan and Han, Bo and Han, Tao},
  booktitle={Proceedings of the 29th Annual International Conference on Mobile Computing and Networking},
  pages={1--15},
  year={2023}
}

@inproceedings{khani2023recl,
  title={$\{$RECL$\}$: Responsive $\{$Resource-Efficient$\}$ continuous learning for video analytics},
  author={Khani, Mehrdad and Ananthanarayanan, Ganesh and Hsieh, Kevin and Jiang, Junchen and Netravali, Ravi and Shu, Yuanchao and Alizadeh, Mohammad and Bahl, Victor},
  booktitle={20th USENIX Symposium on Networked Systems Design and Implementation (NSDI 23)},
  pages={917--932},
  year={2023}
}

@inproceedings{wen2023adaptivenet,
  title={Adaptivenet: Post-deployment neural architecture adaptation for diverse edge environments},
  author={Wen, Hao and Li, Yuanchun and Zhang, Zunshuai and Jiang, Shiqi and Ye, Xiaozhou and Ouyang, Ye and Zhang, Yaqin and Liu, Yunxin},
  booktitle={Proceedings of the 29th Annual International Conference on Mobile Computing and Networking},
  pages={1--17},
  year={2023}
}

@inproceedings{dhawaskar2023converge,
  title={Converge: Qoe-driven multipath video conferencing over webrtc},
  author={Dhawaskar Sathyanarayana, Sandesh and Lee, Kyunghan and Grunwald, Dirk and Ha, Sangtae},
  booktitle={Proceedings of the ACM SIGCOMM 2023 Conference},
  pages={637--653},
  year={2023}
}

@inproceedings{yi2023boosting,
  title={Boosting dnn cold inference on edge devices},
  author={Yi, Rongjie and Cao, Ting and Zhou, Ao and Ma, Xiao and Wang, Shangguang and Xu, Mengwei},
  booktitle={Proceedings of the 21st Annual International Conference on Mobile Systems, Applications and Services},
  pages={516--529},
  year={2023}
}

@inproceedings{zhou2019learning,
  title={Learning to coordinate video codec with transport protocol for mobile video telephony},
  author={Zhou, Anfu and Zhang, Huanhuan and Su, Guangyuan and Wu, Leilei and Ma, Ruoxuan and Meng, Zhen and Zhang, Xinyu and Xie, Xiufeng and Ma, Huadong and Chen, Xiaojiang},
  booktitle={The 25th Annual International Conference on Mobile Computing and Networking},
  pages={1--16},
  year={2019}
}

@article{wang2016algorithmic,
  title={Algorithmic principles of remote PPG},
  author={Wang, Wenjin and Den Brinker, Albertus C and Stuijk, Sander and De Haan, Gerard},
  journal={IEEE Transactions on Biomedical Engineering},
  volume={64},
  number={7},
  pages={1479--1491},
  year={2016},
  publisher={IEEE}
}

@inproceedings{tao2021someone,
  title={Is Someone Speaking? Exploring Long-term Temporal Features for Audio-visual Active Speaker Detection},
  author={Tao, Ruijie and Pan, Zexu and Das, Rohan Kumar and Qian, Xinyuan and Shou, Mike Zheng and Li, Haizhou},
  booktitle = {Proceedings of the 29th ACM International Conference on Multimedia},
  pages = {3927–3935},
  year={2021}
}

@inproceedings{jiang2023target,
  title={Target Active Speaker Detection with Audio-visual Cues},
  author={Jiang, Yidi and Tao, Ruijie and Pan, Zexu and Li, Haizhou},
  booktitle={Proc. Interspeech},
  year={2023}
}

@inproceedings{hu2018squeeze,
  title={Squeeze-and-excitation networks},
  author={Hu, Jie and Shen, Li and Sun, Gang},
  booktitle={Proceedings of the IEEE conference on computer vision and pattern recognition},
  pages={7132--7141},
  year={2018}
}

@article{afouras2018conversation,
  title={The conversation: Deep audio-visual speech enhancement},
  author={Afouras, Triantafyllos and Chung, Joon Son and Zisserman, Andrew},
  journal={arXiv preprint arXiv:1804.04121},
  year={2018}
}

@misc{VideoStreamingSVoD,
	title = {Video {Streaming} ({SVoD}) - {Global} {\textbar} {Statista} {Market} {Forecast}},
	url = {https://www.statista.com/outlook/dmo/digital-media/video-on-demand/video-streaming-svod/worldwide},
	language = {en},
	urldate = {2024-09-05},
	journal = {Statista},
}

@inproceedings{wang2020you,
  title={What you wear know how you feel: An emotion inference system with multi-modal wearable devices},
  author={Wang, Dan and Lei, Haibo and Dong, Haozhi and Wang, Yunshu and Zou, Yongpan and Wu, Kaishun},
  booktitle={Proceedings of the 26th Annual International Conference on Mobile Computing and Networking},
  pages={1--3},
  year={2020}
}

@inproceedings{sun2022estimating,
  title={Estimating stress in online meetings by remote physiological signal and behavioral features},
  author={Sun, Zhaodong and Vedernikov, Alexander and Kykyri, Virpi-Liisa and Pohjola, Mikko and Nokia, Miriam and Li, Xiaobai},
  booktitle={Adjunct Proceedings of the 2022 ACM International Joint Conference on Pervasive and Ubiquitous Computing and the 2022 ACM International Symposium on Wearable Computers},
  pages={216--220},
  year={2022}
}

@inproceedings{liu2019cardiocam,
  title={Cardiocam: Leveraging camera on mobile devices to verify users while their heart is pumping},
  author={Liu, Jian and Shi, Cong and Chen, Yingying and Liu, Hongbo and Gruteser, Marco},
  booktitle={Proceedings of the 17th Annual International Conference on Mobile Systems, Applications, and Services},
  pages={249--261},
  year={2019}
}

@inproceedings{qi2020deeprhythm,
  title={Deeprhythm: Exposing deepfakes with attentional visual heartbeat rhythms},
  author={Qi, Hua and Guo, Qing and Juefei-Xu, Felix and Xie, Xiaofei and Ma, Lei and Feng, Wei and Liu, Yang and Zhao, Jianjun},
  booktitle={Proceedings of the 28th ACM international conference on multimedia},
  pages={4318--4327},
  year={2020}
}

@misc{PolarH10Polar,
	title = {Polar {H10} {\textbar} {Polar} {Global}},
	url = {https://www.polar.com/en/sensors/h10-heart-rate-sensor},
	language = {en},
	urldate = {2024-09-05},
}

@misc{SkypeStayConnected,
	title = {Skype {\textbar} {Stay} connected with free video calls worldwide},
	url = {https://www.skype.com/en//},
	language = {en},
	urldate = {2024-09-05},
}

@misc{TeamsChannelsMicrosoft,
	title = {Teams and {Channels} {\textbar} {Microsoft} {Teams}},
	url = {https://teams.microsoft.com/v2/?clientexperience=t2},
	urldate = {2024-09-05},
}

@misc{OnePlatformConnect,
	title = {One platform to connect {\textbar} {Zoom}},
	url = {https://zoom.us/},
	urldate = {2024-09-05},
}

@misc{StreamTVMovies,
	title = {Stream {TV} and {Movies} {Live} and {Online} {\textbar} {Hulu}},
	url = {https://www.hulu.com/welcome?orig_referrer=https%3A%2F%2Fwww.google.com.hk%2F},
	urldate = {2024-09-05},
}

@misc{NetflixSingaporeWatch,
	title = {Netflix {Singapore} – {Watch} {TV} {Programmes} {Online}, {Watch} {Films} {Online}},
	url = {https://www.netflix.com/sg/},
	urldate = {2024-09-05},
}

@misc{YouTube,
	title = {{YouTube}},
	url = {https://www.youtube.com/},
	urldate = {2024-09-05},
}

@misc{ExploreFindYour,
	title = {Explore - {Find} your favourite videos on {TikTok}},
	url = {https://www.tiktok.com/explore},
	urldate = {2024-09-05},
}

@inproceedings{lion2016don,
  title={$\{$Don’t$\}$ Get Caught in the Cold, Warm-up Your $\{$JVM$\}$: Understand and Eliminate $\{$JVM$\}$ Warm-up Overhead in $\{$Data-Parallel$\}$ Systems},
  author={Lion, David and Chiu, Adrian and Sun, Hailong and Zhuang, Xin and Grcevski, Nikola and Yuan, Ding},
  booktitle={12th USENIX Symposium on Operating Systems Design and Implementation (OSDI 16)},
  pages={383--400},
  year={2016}
}

@article{shafer1985using,
  title={Using color to separate reflection components},
  author={Shafer, Steven A},
  journal={Color Research \& Application},
  volume={10},
  number={4},
  pages={210--218},
  year={1985},
  publisher={Wiley Online Library}
}

@inproceedings{stricker2014non,
  title={Non-contact video-based pulse rate measurement on a mobile service robot},
  author={Stricker, Ronny and M{\"u}ller, Steffen and Gross, Horst-Michael},
  booktitle={The 23rd IEEE International Symposium on Robot and Human Interactive Communication},
  pages={1056--1062},
  year={2014},
  organization={IEEE}
}

@INPROCEEDINGS{10340857,
  author={Tang, Jiankai and Chen, Kequan and Wang, Yuntao and Shi, Yuanchun and Patel, Shwetak and McDuff, Daniel and Liu, Xin},
  booktitle={2023 45th Annual International Conference of the IEEE Engineering in Medicine \& Biology Society (EMBC)}, 
  title={MMPD: Multi-Domain Mobile Video Physiology Dataset}, 
  year={2023},
  volume={},
  number={},
  pages={1-5},
  doi={10.1109/EMBC40787.2023.10340857}}

@inproceedings{long2022dynamic,
  title={Dynamic temporal filtering in video models},
  author={Long, Fuchen and Qiu, Zhaofan and Pan, Yingwei and Yao, Ting and Ngo, Chong-Wah and Mei, Tao},
  booktitle={European Conference on Computer Vision},
  pages={475--492},
  year={2022},
  organization={Springer}
}

@misc{GStreamer,
	title = {{GStreamer}},
	url = {https://gstreamer.freedesktop.org/documentation/?gi-language=c},
	urldate = {2024-09-05},
}

@inproceedings{ravanelli2018speaker,
  title={Speaker recognition from raw waveform with sincnet},
  author={Ravanelli, Mirco and Bengio, Yoshua},
  booktitle={2018 IEEE spoken language technology workshop (SLT)},
  pages={1021--1028},
  year={2018},
  organization={IEEE}
}

@inproceedings{wang2021tdn,
  title={Tdn: Temporal difference networks for efficient action recognition},
  author={Wang, Limin and Tong, Zhan and Ji, Bin and Wu, Gangshan},
  booktitle={Proceedings of the IEEE/CVF conference on computer vision and pattern recognition},
  pages={1895--1904},
  year={2021}
}

@inproceedings{sun2021ultrase,
  title={UltraSE: single-channel speech enhancement using ultrasound},
  author={Sun, Ke and Zhang, Xinyu},
  booktitle={Proceedings of the 27th annual international conference on mobile computing and networking},
  pages={160--173},
  year={2021}
}

@inproceedings{ross2017focal,
  title={Focal loss for dense object detection},
  author={Ross, T-YLPG and Doll{\'a}r, GKHP},
  booktitle={proceedings of the IEEE conference on computer vision and pattern recognition},
  pages={2980--2988},
  year={2017}
}

@inproceedings{mottelson2016affect,
  title={An affect detection technique using mobile commodity sensors in the wild},
  author={Mottelson, Aske and Hornb{\ae}k, Kasper},
  booktitle={Proceedings of the 2016 ACM International Joint Conference on Pervasive and Ubiquitous Computing},
  pages={781--792},
  year={2016}
}

@article{prajwal2023towards,
  title={Towards Efficient Emotion Self-report Collection Using Human-AI Collaboration: A Case Study on Smartphone Keyboard Interaction},
  author={Prajwal, M and Raj, Ayush and Sen, Sougata and Saha, Snehanshu and Ghosh, Surjya},
  journal={Proceedings of the ACM on Interactive, Mobile, Wearable and Ubiquitous Technologies},
  volume={7},
  number={2},
  pages={1--23},
  year={2023},
  publisher={ACM New York, NY, USA}
}

@inproceedings{wu2020emo,
  title={EMO: Real-time emotion recognition from single-eye images for resource-constrained eyewear devices},
  author={Wu, Hao and Feng, Jinghao and Tian, Xuejin and Sun, Edward and Liu, Yunxin and Dong, Bo and Xu, Fengyuan and Zhong, Sheng},
  booktitle={Proceedings of the 18th International Conference on Mobile Systems, Applications, and Services},
  pages={448--461},
  year={2020}
}

@inproceedings{ahmad2024detecting,
  title={Detecting deception in natural environments using incremental transfer learning},
  author={Ahmad, Muneeb Imtiaz and Alzahrani, Abdullah and Ahmad, Sunbul M},
  booktitle={Proceedings of the 26th International Conference on Multimodal Interaction},
  pages={66--75},
  year={2024}
}

@article{yang2023avoid,
  title={Avoid-df: Audio-visual joint learning for detecting deepfake},
  author={Yang, Wenyuan and Zhou, Xiaoyu and Chen, Zhikai and Guo, Bofei and Ba, Zhongjie and Xia, Zhihua and Cao, Xiaochun and Ren, Kui},
  journal={IEEE Transactions on Information Forensics and Security},
  volume={18},
  pages={2015--2029},
  year={2023},
  publisher={IEEE}
}

@inproceedings{demir2021deep,
  title={Where do deep fakes look? synthetic face detection via gaze tracking},
  author={Demir, Ilke and Ciftci, Umur Aybars},
  booktitle={ACM symposium on eye tracking research and applications},
  pages={1--11},
  year={2021}
}

@article{wascher2021heart,
  title={Heart rate as a measure of emotional arousal in evolutionary biology},
  author={Wascher, Claudia AF},
  journal={Philosophical Transactions of the Royal Society B},
  volume={376},
  number={1831},
  pages={20200479},
  year={2021},
  publisher={The Royal Society}
}

@inproceedings{liu2020grad,
  title={Grad: Learning for overhead-aware adaptive video streaming with scalable video coding},
  author={Liu, Yunzhuo and Jiang, Bo and Guo, Tian and Sitaraman, Ramesh K and Towsley, Don and Wang, Xinbing},
  booktitle={Proceedings of the 28th ACM International Conference on Multimedia},
  pages={349--357},
  year={2020}
}

@article{mmdetection,
  title   = {{MMDetection}: Open MMLab Detection Toolbox and Benchmark},
  author  = {Chen, Kai and Wang, Jiaqi and Pang, Jiangmiao and Cao, Yuhang and
             Xiong, Yu and Li, Xiaoxiao and Sun, Shuyang and Feng, Wansen and
             Liu, Ziwei and Xu, Jiarui and Zhang, Zheng and Cheng, Dazhi and
             Zhu, Chenchen and Cheng, Tianheng and Zhao, Qijie and Li, Buyu and
             Lu, Xin and Zhu, Rui and Wu, Yue and Dai, Jifeng and Wang, Jingdong
             and Shi, Jianping and Ouyang, Wanli and Loy, Chen Change and Lin, Dahua},
  journal= {arXiv preprint arXiv:1906.07155},
  year={2019}
}

@inproceedings{du2020server,
  title={Server-driven video streaming for deep learning inference},
  author={Du, Kuntai and Pervaiz, Ahsan and Yuan, Xin and Chowdhery, Aakanksha and Zhang, Qizheng and Hoffmann, Henry and Jiang, Junchen},
  booktitle={Proceedings of the Annual conference of the ACM Special Interest Group on Data Communication on the applications, technologies, architectures, and protocols for computer communication},
  pages={557--570},
  year={2020}
}

@article{wiseman2016python,
  title={Python interface to the webrtc voice activity detector},
  author={Wiseman, John and Bondarenko, Ivan Yu},
  journal={Python interface to the WebRTC voice activity detector},
  year={2016}
}

@article{yang2022survey,
  title={Survey on emotion sensing using mobile devices},
  author={Yang, Kangning and Tag, Benjamin and Wang, Chaofan and Gu, Yue and Sarsenbayeva, Zhanna and Dingler, Tilman and Wadley, Greg and Goncalves, Jorge},
  journal={IEEE Transactions on Affective Computing},
  volume={14},
  number={4},
  pages={2678--2696},
  year={2022},
  publisher={IEEE}
}

@article{fairclough2020personal,
  title={Personal informatics and negative emotions during commuter driving: Effects of data visualization on cardiovascular reactivity \& mood},
  author={Fairclough, Stephen H and Dobbins, Chelsea},
  journal={International Journal of Human-Computer Studies},
  volume={144},
  pages={102499},
  year={2020},
  publisher={Elsevier}
}

@article{bian2024ubihr,
  title={UbiHR: Resource-efficient Long-range Heart Rate Sensing on Ubiquitous Devices},
  author={Bian, Haoyu and Guo, Bin and Liu, Sicong and Ding, Yasan and Gao, Shanshan and Yu, Zhiwen},
  journal={Proceedings of the ACM on Interactive, Mobile, Wearable and Ubiquitous Technologies},
  volume={8},
  number={4},
  pages={1--26},
  year={2024},
  publisher={ACM New York, NY, USA}
}

@inproceedings{zhang2024aralive,
  title={AraLive: Automatic Reward Adaption for Learning-based Live Video Streaming},
  author={Zhang, Huanhuan and zhuo, Liu and Li, Haotian and Zhou, Anfu and Wang, Chuanming and Ma, Huadong},
  booktitle={Proceedings of the 32nd ACM International Conference on Multimedia},
  pages={11099--11108},
  year={2024}
}

@inproceedings{wu2024toward,
  title={Toward Timeliness-Enhanced Loss Recovery for Large-Scale Live Streaming},
  author={Wu, Bo and Li, Tong and Luo, Cheng and Yan, Xu and Wang, Fuyu and Du, Xinle and Xu, Ke},
  booktitle={Proceedings of the 32nd ACM International Conference on Multimedia},
  pages={7891--7899},
  year={2024}
}

@inproceedings{qian2024cluster,
  title={Cluster-phys: Facial clues clustering towards efficient remote physiological measurement},
  author={Qian, Wei and Li, Kun and Guo, Dan and Hu, Bin and Wang, Meng},
  booktitle={Proceedings of the 32nd ACM International Conference on Multimedia},
  pages={330--339},
  year={2024}
}

@inproceedings{wang2024rppg,
  title={rPPG-HiBa: Hierarchical Balanced Framework for Remote Physiological Measurement},
  author={Wang, Yin and Lu, Hao and Chen, Ying-Cong and Kuang, Li and Zhou, Mengchu and Deng, Shuiguang},
  booktitle={Proceedings of the 32nd ACM International Conference on Multimedia},
  pages={2982--2991},
  year={2024}
}

@inproceedings{xue2025ppg,
  title={PPG Earring: Wireless Smart Earring for Heart Health Monitoring},
  author={Xue, Qiuyue and Nissanka, Dilini and Yan, Jiachen Tammy and Wang, Ruiqing and Patel, Shwetak and Iyer, Vikram},
  booktitle={Proceedings of the 2025 CHI Conference on Human Factors in Computing Systems},
  pages={1--16},
  year={2025}
}

@article{lyu2024ase,
  title={ASE: Practical Acoustic Speed Estimation Beyond Doppler via Sound Diffusion Field},
  author={Lyu, Sheng and Wu, Chenshu},
  journal={arXiv preprint arXiv:2412.20142},
  year={2024}
}

@inproceedings{xue2025ecg,
  title={ECG Necklace: Low-power Wireless Necklace for Continuous ECG monitoring},
  author={Xue, Qiuyue and Martin, Eric Steven and Liu, Jiaqing and Wang, Ruiqing and Glenn, Antonio and Li, Richard and Iyer, Vikram and Patel, Shwetak},
  booktitle={Proceedings of the 2025 CHI Conference on Human Factors in Computing Systems},
  pages={1--14},
  year={2025}
}

@inproceedings{cao2024earsteth,
  title={EarSteth: Cardiac Auscultation Audio Reconstruction Using Earbuds},
  author={Cao, Alvin and Christofferson, Ken and Ruth, Parker and Rabbani, Naveed and Shi, Yuanchun and Mariakakis, Alex and Wang, Yuntao and Patel, Shwetak},
  booktitle={2024 46th Annual International Conference of the IEEE Engineering in Medicine and Biology Society (EMBC)},
  pages={1--4},
  year={2024},
  organization={IEEE}
}

@inproceedings{li2022livenet,
  title={Livenet: a low-latency video transport network for large-scale live streaming},
  author={Li, Jinyang and Li, Zhenyu and Lu, Ri and Xiao, Kai and Li, Songlin and Chen, Jufeng and Yang, Jingyu and Zong, Chunli and Chen, Aiyun and Wu, Qinghua and others},
  booktitle={Proceedings of the ACM SIGCOMM 2022 Conference},
  pages={812--825},
  year={2022}
}

@inproceedings{chen2024videollm,
  title={Videollm-online: Online video large language model for streaming video},
  author={Chen, Joya and Lv, Zhaoyang and Wu, Shiwei and Lin, Kevin Qinghong and Song, Chenan and Gao, Difei and Liu, Jia-Wei and Gao, Ziteng and Mao, Dongxing and Shou, Mike Zheng},
  booktitle={Proceedings of the IEEE/CVF Conference on Computer Vision and Pattern Recognition},
  pages={18407--18418},
  year={2024}
}

@article{qian2024streaming,
  title={Streaming long video understanding with large language models},
  author={Qian, Rui and Dong, Xiaoyi and Zhang, Pan and Zang, Yuhang and Ding, Shuangrui and Lin, Dahua and Wang, Jiaqi},
  journal={Advances in Neural Information Processing Systems},
  volume={37},
  pages={119336--119360},
  year={2024}
}

@article{li2025videoscan,
  title={VideoScan: Enabling Efficient Streaming Video Understanding via Frame-level Semantic Carriers},
  author={Li, Ruanjun and Tan, Yuedong and Shi, Yuanming and Shao, Jiawei},
  journal={arXiv preprint arXiv:2503.09387},
  year={2025}
}

@article{tong2022videomae,
  title={Videomae: Masked autoencoders are data-efficient learners for self-supervised video pre-training},
  author={Tong, Zhan and Song, Yibing and Wang, Jue and Wang, Limin},
  journal={Advances in neural information processing systems},
  volume={35},
  pages={10078--10093},
  year={2022}
}

@article{naiman2025lv,
  title={LV-MAE: Learning Long Video Representations through Masked-Embedding Autoencoders},
  author={Naiman, Ilan and Ben-Baruch, Emanuel and Anschel, Oron and Shoshan, Alon and Kviatkovsky, Igor and Aggarwal, Manoj and Medioni, Gerard},
  journal={arXiv preprint arXiv:2504.03501},
  year={2025}
}

@inproceedings{zou2025rhythmmamba,
  title={RhythmMamba: Fast, Lightweight, and Accurate Remote Physiological Measurement},
  author={Zou, Bochao and Guo, Zizheng and Hu, Xiaocheng and Ma, Huimin},
  booktitle={Proceedings of the AAAI Conference on Artificial Intelligence},
  volume={39},
  number={10},
  pages={11077--11085},
  year={2025}
}

@inproceedings{luo2024physmamba,
  title={PhysMamba: Efficient remote physiological measurement with SlowFast temporal difference mamba},
  author={Luo, Chaoqi and Xie, Yiping and Yu, Zitong},
  booktitle={Chinese Conference on Biometric Recognition},
  pages={248--259},
  year={2024},
  organization={Springer}
}

@article{wu2025cardiacmamba,
  title={CardiacMamba: A Multimodal RGB-RF Fusion Framework with State Space Models for Remote Physiological Measurement},
  author={Wu, Zheng and Xie, Yiping and Zhao, Bo and He, Jiguang and Luo, Fei and Deng, Ning and Yu, Zitong},
  journal={arXiv preprint arXiv:2502.13624},
  year={2025}
}

@inproceedings{tashtarian2024artemis,
  title={$\{$ARTEMIS$\}$: adaptive bitrate ladder optimization for live video streaming},
  author={Tashtarian, Farzad and Bentaleb, Abdelhak and Amirpour, Hadi and Gorinsky, Sergey and Jiang, Junchen and Hellwagner, Hermann and Timmerer, Christian},
  booktitle={21st USENIX Symposium on Networked Systems Design and Implementation (NSDI 24)},
  pages={591--611},
  year={2024}
}

@inproceedings{bui2019ebp,
  title={ebp: A wearable system for frequent and comfortable blood pressure monitoring from user's ear},
  author={Bui, Nam and Pham, Nhat and Barnitz, Jessica Jacqueline and Zou, Zhanan and Nguyen, Phuc and Truong, Hoang and Kim, Taeho and Farrow, Nicholas and Nguyen, Anh and Xiao, Jianliang and others},
  booktitle={The 25th annual international conference on mobile computing and networking},
  pages={1--17},
  year={2019}
}

@article{alimbayeva2022portable,
  title={Portable ECG Monitoring System},
  author={Alimbayeva, Zhadyra N and Alimbayev, Chingiz A and Bayanbay, Nurlan A and Ozhikenov, Kassymbek A and Bodin, Oleg N and Mukazhanov, Yerkat B},
  journal={International Journal of Advanced Computer Science and Applications},
  volume={13},
  number={4},
  year={2022},
  publisher={Science and Information (SAI) Organization Limited}
}

@article{chan2025mobile,
  title={Mobile medical systems for equitable healthcare},
  author={Chan, Justin and Goel, Mayank and Gollakota, Shyamnath and Nandakumar, Rajalakshmi},
  journal={Nature Reviews Bioengineering},
  pages={1--20},
  year={2025},
  publisher={Nature Publishing Group UK London}
}

@article{park2020heartquake,
  title={Heartquake: Accurate low-cost non-invasive ecg monitoring using bed-mounted geophones},
  author={Park, Jaeyeon and Cho, Hyeon and Balan, Rajesh Krishna and Ko, JeongGil},
  journal={Proceedings of the ACM on Interactive, Mobile, Wearable and Ubiquitous Technologies},
  volume={4},
  number={3},
  pages={1--28},
  year={2020},
  publisher={ACM New York, NY, USA}
}

@article{chan2019contactless,
  title={Contactless cardiac arrest detection using smart devices},
  author={Chan, Justin and Rea, Thomas and Gollakota, Shyamnath and Sunshine, Jacob E},
  journal={NPJ digital medicine},
  volume={2},
  number={1},
  pages={52},
  year={2019},
  publisher={Nature Publishing Group UK London}
}

@article{gong2022contrastive,
  title={Contrastive audio-visual masked autoencoder},
  author={Gong, Yuan and Rouditchenko, Andrew and Liu, Alexander H and Harwath, David and Karlinsky, Leonid and Kuehne, Hilde and Glass, James},
  journal={arXiv preprint arXiv:2210.07839},
  year={2022}
}

@inproceedings{zhang2016multimodal,
  title={Multimodal spontaneous emotion corpus for human behavior analysis},
  author={Zhang, Zheng and Girard, Jeff M and Wu, Yue and Zhang, Xing and Liu, Peng and Ciftci, Umur and Canavan, Shaun and Reale, Michael and Horowitz, Andy and Yang, Huiyuan and others},
  booktitle={Proceedings of the IEEE conference on computer vision and pattern recognition},
  pages={3438--3446},
  year={2016}
}

\end{document}